\documentclass[showpacs,preprintnumbers,amsmath,amssymb]{revtex4}
\usepackage{epsfig}
\usepackage{graphicx}
\begin{document}

\title[Magnetoresistance and dephasing in a 2D electron gas
at intermediate conductances] {Magnetoresistance and dephasing in
a two-dimensional electron gas at intermediate conductances }

\author{G.~M.~Minkov}
\author{A.~V.~Germanenko}

\affiliation{Institute of Physics and Applied Mathematics, Ural
State University, 620083 Ekaterinburg, Russia}

\author{I.~V.~Gornyi$^*$}

\affiliation{Institut f\"{u}r Nanotechnologie, Forschungszentrum
Karlsruhe, 76021 Karlsruhe, Germany}

\date{\today}

\begin{abstract}

We study, both theoretically and experimentally, the negative
magnetoresistance (MR) of a two-dimensional (2D) electron gas in a weak transverse
magnetic field $B$.
The analysis is carried out in a wide range of zero-$B$ conductances $g$
(measured in units of $e^2/h$),
including the range of intermediate conductances, $g\sim 1$.
Interpretation of the experimental results obtained for a
2D electron gas in GaAs/In$_x$Ga$_{1-x}$As/GaAs single quantum
well structures is based on the theory which takes into account terms of
higher orders in $1/g$. We show that the standard
weak localization (WL) theory is adequate for $g\gtrsim 5$.
Calculating the corrections of second order in $1/g$ to the MR,
stemming from both the interference contribution and the
mutual effect of WL and Coulomb interaction,
we expand the range of a quantitative agreement
between the theory and experiment down to significantly lower
conductances $g\sim 1$.
We demonstrate that at intermediate conductances the
negative MR is described
by the standard WL ``digamma-functions'' expression,
but with a reduced prefactor $\alpha$.
We also show that at not very high $g$
the second-loop corrections dominate over the contribution
of the interaction in the Cooper channel,
and therefore appear to be the main source of the lowering of
the prefactor, $\alpha\simeq 1-2/\pi g$.
The fitting of the MR allows us to measure
the true value of the phase breaking time within a wide
conductance range, $g\gtrsim 1$.
We further analyze the regime of a ``weak insulator'', when the zero-$B$
conductance is low $g(B=0)<1$ due to the localization at low temperature,
whereas the Drude conductance is high, $g_0\gg 1,$ so that a weak
magnetic field delocalizes electronic states. In this regime,
while the MR still can be fitted by the digamma-functions formula,
the experimentally obtained value of the dephasing rate
has nothing to do with the true one.
The corresponding fitting parameter in the low-$T$ limit
is determined by the localization length and may therefore
saturate at $T\to 0$, even though the true dephasing rate
vanishes.

\end{abstract}


 \maketitle

\section{Introduction}
\label{sec:int}

Conventional theories of weak localization (WL) and interaction
corrections to the conductivity (for review see
Refs.~\onlinecite{Altshuler,AAKL-MIR,LeeRam,fink,aleiner}) are
developed for the case $k_F l \gg 1$, where $k_F$ and $l$ are the
Fermi quasimomentum and the classical mean free path,
respectively. They are valid when the quantum corrections are
small in magnitude compared with the Drude conductivity
\begin{equation}
\sigma_0=\frac{e^2n\tau}{m}=2 e^2 \nu D =\pi k_F l\, G_0, \label{eq1}
\end{equation}
where $n$ and $m$ denote electron density and mass, respectively,
$\tau$ is the elastic transport mean free time, $D$ is the
diffusion constant, $\nu=m/2\pi \hbar^2$ is the density of states
per spin, and $G_0=e^2/(2\pi^2\hbar)$. In two-dimensional (2D)
systems the quantum corrections arising due to interference and/or
interaction effects are logarithmic in temperature $T$ at low
temperatures.

The situation when $k_F l \gg 1$ and the quantum corrections are
comparable in magnitude to the Drude conductivity is quite
unrealistic. For example, using the well-known expressions for the
phase relaxation time~\cite{Altshuler,AAKL-MIR} $\tau_\varphi$ one
can easily find that the interference
correction~\cite{GorkovLarkinKhmelnitskii},
$\delta\sigma_{WL}=-G_0\ln(\tau_\varphi/\tau),$ is less than 15\%
of the Drude conductivity even for $T=10$ mK, when considering the
2D electron gas in GaAs with $n=4\times 10^{15}$~m$^{-2}$ and $k_F
l=20$. Therefore, for high values of the dimensionless conductance
$g_0\equiv k_F l$, the conventional WL theory works perfectly down
to very low temperatures. In reality, the situation when
$\delta\sigma$ and $\sigma_0$ are of the same magnitude occurs at
$k_F l\simeq 2-5$. In this case the corrections to the
conductivity of higher orders in $(k_F l)^{-1}$ become important
and the WL theory is not expected to work. This range of
intermediate conductances is addressed in the present paper.

Fundamentally, the properties of 2D systems are
controlled by several characteristic length scales. At zero
temperature in two dimensions
the disordered wave function is always localized~\cite{gang4}
over the length scale $\xi,$ which can be estimated
as~\cite{LeeRam}
\begin{equation}
\xi=\xi_O\simeq l \exp(\pi k_F l/2). \label{eq:XiO}
\end{equation}
Here the subscript $O$ refers to the orthogonal
symmetry of the disordered Hamiltonian.
Real experiments are carried out at nonzero temperature and
another length scale $L_\varphi=(D\tau_\varphi)^{1/2},$
over which electrons maintain phase coherence, arises in this case.
In semiconductor 2D systems, at low
temperatures the inelasticity of electron-electron
interaction is the main source of the phase breaking
processes~\cite{Altshuler,AAK82}
and $\tau_\varphi \propto T^{-1}.$

\begin{figure}
\includegraphics[width=0.6\linewidth,clip=true]{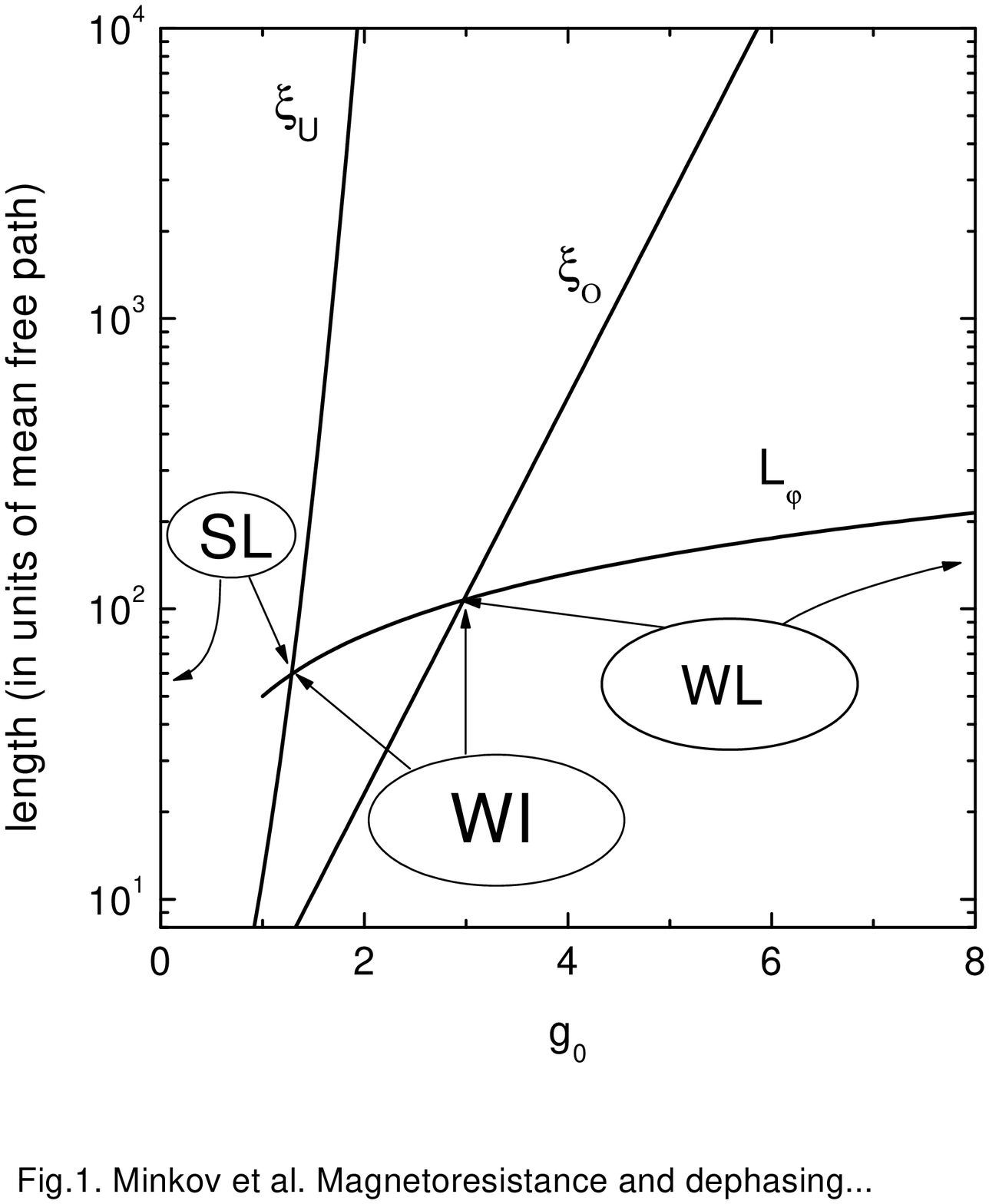}
 \caption
{Schematic representation of the characteristic scale lengths
$L_\varphi$, $\xi_O$, and $\xi_U$ plotted versus conductance
$g_0\equiv k_Fl$.} \label{f0}
\end{figure}

Measurement of the magnetoresistance (MR) is one of the most
useful tools for investigation of physical properties of a 2D electron gas.
An external transverse magnetic field $B$ destroys the quantum interference and
therefore influences the localization. It breaks the time reversal
invariance, thus changing the symmetry of the disordered
Hamiltonian from orthogonal to unitary. As a result, the localization
length becomes $B$-dependent~\cite{Lerner-Imry}, $\xi=\xi(B),$ and
changes with increasing $B$ from $\xi_O$ to $\xi_U$.
The latter for classically weak
magnetic fields can be estimated as~\cite{Hik81}
\begin{equation}
\xi_U\simeq l \exp\left[(\pi k_F l/2)^2\right], \label{eq:XiU}
\end{equation}
that is much greater than $\xi_O$ for $k_F l\gg 1$. Thus, there
are three key length scales $\xi_O$, $\xi_U,$ and $L_\varphi,$
that determine the state of a 2D system and its transport
properties. In Fig.~\ref{f0} we illustrate schematically the
behavior of the lengths $\xi_O$, $\xi_U$, and $L_\varphi$ with
changing the conductance $g_0$ at a given temperature. In what
follows, we will consider the case $k_F l>1$.

When the phase breaking length is
much shorter than the localization lengths, $L_\varphi\ll\xi_O,\,\xi_U$,
the system is in the WL regime
for an arbitrary magnetic field.
In classically strong magnetic fields, the MR
is produced by the interaction-induced Altshuler-Aronov
correction to the conductivity (see Ref.~\onlinecite{GornyiMirlin} for
review). At low magnetic fields, the negative MR, arising due to
the suppression of quantum interference, is a well-known
manifestation of weak localization~\cite{Altshuler,Hik,AKhLL}.
This effect will be the subject of the present paper.

In Fig.~\ref{f0}, for values of $g_0$ lying
to the left from the point of intersection
of $\xi_U$ and $L_\varphi$ curves~\cite{foot-Lvstau},
the 2D system is in a strong localization (SL) regime.
It is commonly believed that the
transport in the SL case is of a hopping~\cite{SE} nature.
The magnetoresistance in this
regime is also related to the influence of magnetic field on the
quantum interference and has been studied in
Refs.~\onlinecite{spivak,Sarachik,Imry1,Imry2}. In particular, a
parabolic low-field MR in the hopping regime was predicted.
However, the results based on the conventional hopping picture
cannot be directly applied to the experimental
situation addressed in this paper, even when $\sigma(B=0)\lesssim G_0$.
This is because the usual concepts of
hopping (including the percolation treatment)
are justified only when the disorder is large.

Finally, there is an intermediate regime which we term
{\it ``weak insulator''} (WI) regime, when
the phase-breaking length $L_\varphi$ is between the $\xi_U$ and $\xi_O$
lines.
In this regime electrons are localized at zero $B$ and the WL
theory does not work. However, already a very weak magnetic field
shifts the actual localization length $\xi(B)$ toward $\xi_U$
making $\xi(B)>L_\varphi.$
In such a situation the transport is again of the diffusive nature.
Therefore, the theory of weak localization can be applied when the
MR is considered for $\xi_O\ll L_\varphi \ll
\xi(B)$, even though at zero $B$ the total conductance is smaller
than unity. Obviously, this situation is only possible when $k_F
l\gtrsim 1$. This is a necessary condition for opening
a window between the two localization lengths $\xi_O$ and $\xi_U$.
The WI-problem with a low conductance $g(T) < e^2/h$ at $B=0$ but
with $g_0=k_F l>1$ should be therefore contrasted with the
conventional SL problem with $k_F l \ll 1$ where the conductivity
mechanism is the hopping. It is worth mentioning, however, that
in three-dimensional systems near the mobility edge, a magnetic
field also leads to the delocalization of electronic
states (reentrance phenomenon),
giving rise to a shift of the mobility edge, in
a close similarity to the 2D WI regime~\cite{3DWI}.
Actually, the WI-regime has also much in common with the
notion of a ``moderate insulator'', introduced in
Ref.~\onlinecite{gershenson-1D} to describe the crossover
between the WL and SL regimes in  quasi-one-dimensional
semiconductor wires.

Generally speaking, at $k_F l \gtrsim 1$ the nature
of transport of interacting electrons
at very low $T$ (when the states are
localized with large enough localization length $\xi_O\gg l$) is
not fully understood. At
microscopic scales, $l<L<\xi_O,$ the electron dynamics is diffusive. The
magnetic field serves as a probe of these scales and hence the
MR can provide important information about
the crossover between the localization and diffusion. The WL
theory can be generalized (using, e.g., scaling arguments) to
describe this crossover. Of course, the corrections of higher
orders in $(k_F l)^{-1}$ become then important. Thus it is
desirable to understand the role of these corrections in
magnetotransport.

Apart from the scaling theory of Anderson localization,
there is a self-consistent theory~\cite{Woelfle,Vollhardt-Woelfle},
which enables to calculate
the conductivity for arbitrary value of the quantum interference
correction for $B=0$. However, as we will show below, while at
zero $B$ the self-consistent theory works rather
well, its generalization to the case of finite $B$
fails to describe correctly the magnetoconductivity (MC)
(see Ref.~\onlinecite{Vollhardt-Woelfle} for discussion)
in the crossover between the diffusive and localized regimes.
In this paper we will concentrate on study of higher-order corrections
to the conductivity
using a systematic perturbation theory and scaling approach.

Experimentally, the low-field magnetoresistance in the range of
intermediate conductances and in the crossover regime between the
diffusion and localization in 2D systems has been studied in
Refs.~\onlinecite{Pepper,Davies83,Hsu,Simmons00,proskur,
brunthaler,WLtoSL,coleridge, pusep,krav}. It turns out that the MC
even at low conductance, $\sigma(B=0)\lesssim G_0$, can be still
fitted by the well-known weak-localization
expression~\cite{Hik,AKhLL} derived for $k_F l \gg 1$ (we will
term it the WLMC-formula throughout the paper), but with a reduced
prefactor $\alpha<1$~\cite{WLtoSL}. Similar observations have been
recently reported in Refs.~\onlinecite{pusep,krav}. In
Refs.~\onlinecite{pusep} the magnetotransport has been studied in
quasi-2D systems (doped GaAs/AlGaAs superlattices) and the MR has
been shown to be generated by the quantum interference. A
self-consistent theory of the MC has been employed to fit the
data. In Ref.~\onlinecite{krav}, the MR in a weak perpendicular
magnetic field was measured in the vicinity of an apparent
metal-insulator transition~\cite{AKS01,AMP01} in a $Si$ structure
of $n$-type. In this experiment, the magnetoresistance on the
metallic side was perfectly fitted by the WL formula with the
prefactor $\alpha$ decreasing with lowering the density (i.e. upon
approaching the transition). At the lowest density, the value
$\alpha$ was reported~\cite{krav} to be $\sim 10$ times smaller
than that obtained deeply in the metallic state. Finally, the
authors of Ref.~\onlinecite{Hsu}, who measured the
magnetoconductance in ultrathin metallic films, claimed that while
for $\sigma(B=0)>G_0$ the MR is well described by the WL formula,
for $\sigma(B=0)<G_0$ the MR corresponds to the hopping picture.

An important quantity extracted from the measured low-field MR is
the phase breaking time $\tau_\varphi$, usually treated as a
fitting parameter in the WLMC-formula. With the decreasing of the
conductance, the corrections to this formula become more
pronounced and thus the extracted value of $\tau_\varphi$ may
strongly deviate from the true one. Therefore, there is a clear
need for a systematic (both theoretical and experimental) analysis
of the MR at decreasing conductance, including the crossover
regime $\sigma(B=0)\sim G_0.$ A large scatter of experimental data
on the phase breaking time which is evident even in the case $k_F
l\gg 1$ renders reliable interpretation of the data at
intermediate values of $k_Fl$ difficult. Some of these reasons
have been understood. These are the influence of $\delta$-doped
layers~\cite{Tau-phi}, dynamical defects~\cite{dyndef}, and
macroscopic inhomogeneities~\cite{macro} on the phase relaxation
time, the temperature dependence of the mobility of electrons in
quantum well due to temperature dependent disorder in the doped
layers~\cite{e-e}, and the scattering on magnetic
impurities~\cite{VG}. Nevertheless, the results obtained in
Ref.~\onlinecite{WLtoSL} for both interference and
electron-electron contributions to the conductivity in the range
of not very high values of $g_0$ are (surprisingly) in a
qualitative agreement with the existing theories of conductivity
corrections, developed for high conductance. It was shown in
Ref.~\onlinecite{WLtoSL} that at not very high values of $k_Fl$
(at low electron densities), the role of the interaction
correction to the conductivity becomes less important and the main
effect comes from the interference. (This is because the
interaction correction in the triplet
channel~\cite{Altshuler,fink} increases with decreasing $k_F$, and
tends to cancel out the exchange contribution.) However, the
experimental results have been interpreted in
Ref.~\onlinecite{WLtoSL} only qualitatively.

In this paper we present the results of a quantitative analysis of
the interference corrections to the conductivity and the negative MR
at decreasing $k_F l$. We are not going to discuss a theory of
the MR in the range $k_Fl < 1$ and $\sigma(B\neq 0)\ll G_0,$
corresponding to the SL regime. On the other hand, we
address, in particular, the WI regime, when the zero-B conductivity can be
less than $G_0$ at low $T$.

The interpretation of experimental results obtained for
2D electron gas in GaAs/In$_x$Ga$_{1-x}$As/GaAs single quantum
well structures is based on the theory taking into account terms of
higher order in $(k_F l)^{-1}$. We show that the standard ``one-loop''
WL theory is adequate for $\sigma\gtrsim (10-20)\,G_0$.
Calculating corrections of
the next (``second-loop'') order, we expand the range of the quantitative agreement
between the theory and experiments down to significantly lower
conductivity of about $3\,G_0$. This is largely related to a fortunate
circumstance that ${\cal O}(1/g^3)$-terms are absent in the
perturbative expansion of beta-functions
governing the scaling of the conductance~\cite{Hik81}. Therefore, the corrections
to the second-loop expressions derived in this paper
are proportional to $(G_0/\sigma)^2$ and hence turn out to be
numerically small at such values of the conductivity~\cite{foot-nonpert}.

We demonstrate that the WLMC-formula can be still used to fit the
MR in the crossover from the WL and WI regimes. It is shown that
the main effect of higher-order terms is a reduction of the
prefactor $\alpha$ in the these formulas,
\begin{equation}
\alpha \simeq 1-\frac{2 G_0}{\sigma}.
\label{alpha-vs-sigma}
\end{equation}
This expression appears to be applicable for $\alpha\gtrsim 0.3,$ when the fitting
procedure is carried out in a broad range of magnetic fields.
Thus, it becomes possible to
experimentally determine the phase breaking time within a wide conductivity
range, $\sigma\simeq(3-60) G_0$. Moreover,
the qualitative agreement between the experimental data and the
(properly modified) WL theory persists down to significantly smaller
zero-$B$ conductivity $\sigma(T,B=0)\lesssim G_0$, provided that $k_F l > 1$.
In other words, one of the main results of this paper is that the
theory of quantum corrections to the conductivity works rather
well at the limit of its applicability,
i.e. even for ``intermediate'' values of $g$ of order of unity,
down to $\sigma\sim e^2/h$.

We also show that when applied to the MC in the WI regime,
$\xi_O<L_\varphi<\xi_U$, the fitting procedure based on the
conventional WLMC-formula, yields the value of the dephasing rate
which deviates from the real one and contains information about
the localization length, $\xi_O.$ This observation may be relevant
to the explanation of the tendency to a low-$T$ saturation of the
experimentally extracted dephasing time reported recently in
Refs.~\onlinecite{proskur,brunthaler}, where WL was studied at
intermediate conductances in the vicinity of the apparent
metal-insulator transition.

The paper is organized as follows. The next three sections are
devoted to a theoretical consideration of the problem of the
dephasing and quantum corrections to the conductivity. In
Section~\ref{sec:th}, we recall the basic theoretical results on
the dephasing, interference correction, and interference induced
negative MR. Primary emphasis is put on the possible reasons of
the above mentioned fact that the low-field negative MR is
practically always well described by the WLMC-expression with the
reduced prefactor $\alpha$. In Section~\ref{sec:21} we take a
close look at the interaction correction in the Cooper channel,
which is most frequently invoked for the explanation of the
reduction of $\alpha$. The theory of interference quantum
corrections developed in the next order in $1/g$ is expounded in
Section~\ref{sec:3}. The experimental results and their analysis
are presented in Sections~\ref{sec:exp} and~\ref{sec:exp1} .
Finally,  Section~\ref{sec:concl} is devoted to the conclusions.

\section{Dephasing, interference correction, and magnetoconductivity}
\label{sec:th}
 \subsection{Dephasing time in zero magnetic field}
\label{ssec:1}

Let us start with the consideration of WL effects
in zero magnetic field at large values of the conductance, $g\gg
1$ (here the conductance is measured in units of $\pi G_0$).
This condition allows one to treat the dynamics of a particle
quasiclassically, relating the conductivity correction
to the return probability. Within
the framework of the conventional theory of the WL
developed in the first order in $1/g$, the interference quantum
correction in a 2D system
is given by~\cite{Altshuler,AAKL-MIR,LeeRam,aleiner,GorkovLarkinKhmelnitskii}
\begin{equation}
\frac{\delta\sigma}{G_0}=-\ln\left(\frac{\tau_\varphi}{\tau}\right).
\label{eq10}
\end{equation}
This result holds within the diffusion approximation, justified
for $\tau_\varphi/\tau\gg 1$.
In this paper, we will restrict
ourselves to the diffusive regime [$\tau_\varphi,\hbar/(k_B T)\gg
\tau$] and will not consider the ballistic contribution.

In the WL theory, the phase breaking
(also known as phase relaxation, dephasing, or
decoherence) time $\tau_\varphi$ is a characteristic time scale at
which the two waves traversing along the same path in opposite
directions lose their relative phase coherence due to inelastic
scattering events. At longer times (or trajectories' lengths) the
two waves do not interfere and therefore do not contribute to the
WL correction to the conductivity. At low
temperatures the main source of the inelastic scattering is the
Coulomb electron--electron (e-e) interaction.
In this paper, we will
not address the contribution of other decoherence
mechanisms such as electron--phonon interactions, scattering on
dynamical defects, interaction with magnetic impurities, etc.
Generally, the phase relaxation
time is different from other inelastic scattering times, e.g.
the energy-relaxation time~\cite{AAK82,Altshuler,Aleiner2002}.
Moreover, the phase relaxation may depend on the geometry of the
system. In particular, the damping of Aharonov-Bohm oscillations
in quasi-one-dimensional rings differs~\cite{ludwig-mirlin} from the phase
relaxation rate found for infinite wires~\cite{AAK82}.
It is worth mentioning, however, that the same conventional WL phase
relaxation time governs the temperature behavior of mesoscopic
conductance fluctuations~\cite{Aleiner2002} and the (two-loop)
WL correction in the unitary ensemble~\cite{polyakov-samokhin}.

\begin{figure}
\includegraphics[width=0.7\linewidth,clip=true]{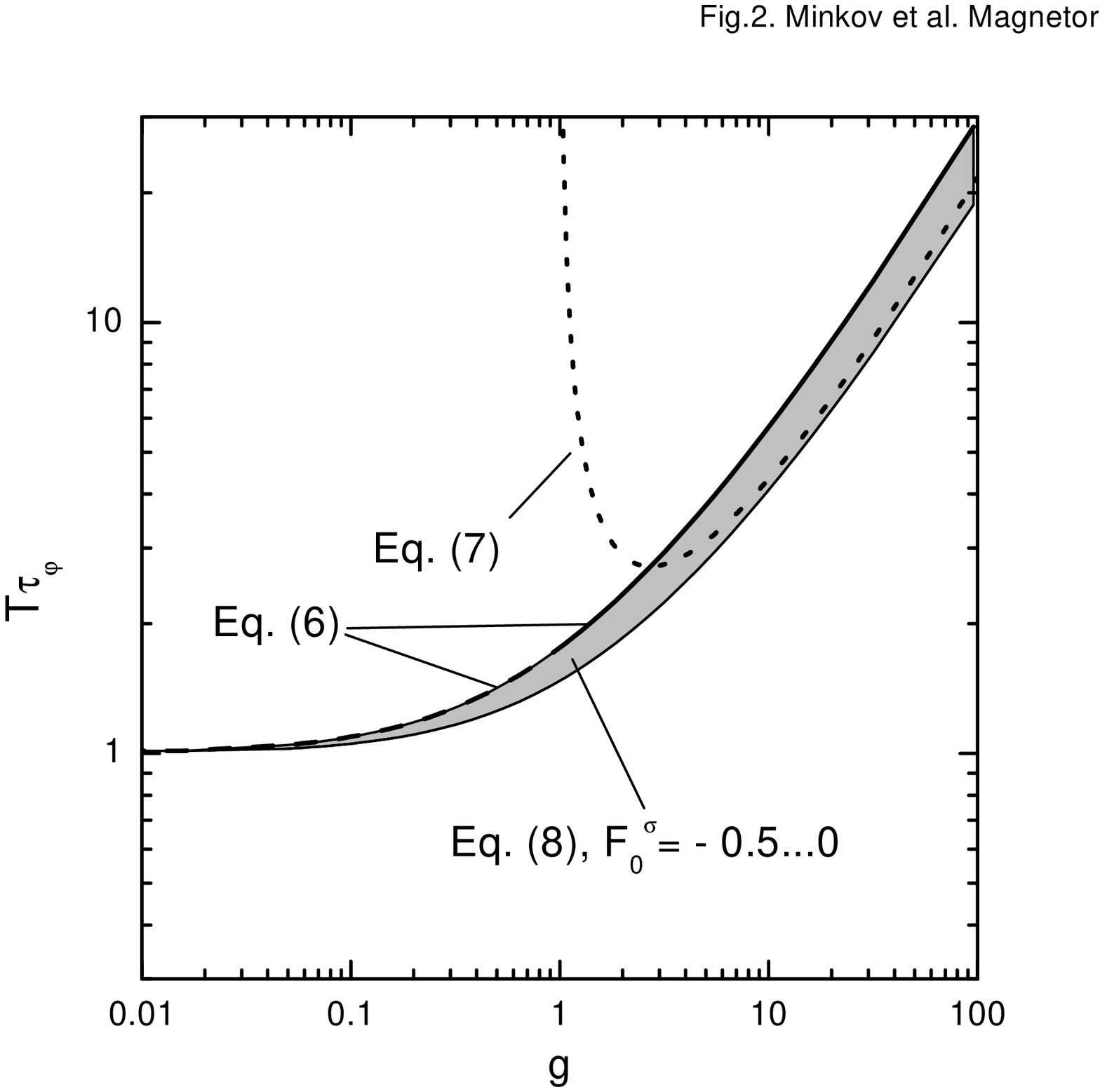}
 \caption
{The conductivity dependence of $T \tau_\varphi$. Solid line is
the solution of Eq.~(\ref{eq11}) (which coincides with
Eq.~(\ref{eq13}) with $F_0^\sigma=0$),  dotted line is the first
iteration Eq.~(\ref{eq109}) for Eq.~(\ref{eq11}). The formal
solution of Eq.~(\ref{eq11}) for $g<1$ is shown by the dashed
curve. Shadow area represents the solutions of Eq.~(\ref{eq13})
found numerically for different values of $F_0^\sigma$ from the
range $-0.5...0$. We set $k_B=\hbar=1$ here.} \label{fig1}
\end{figure}

The phase breaking rate $\tau_\varphi^{-1}$ can be calculated using
the path-integral approach and/or
perturbative diagrammatics~\cite{Altshuler,AAK82,Eiler,aleiner,Aleiner2002}.
As was shown
in Ref.~\onlinecite{AAK82}, the inelastic scattering events with energy
transfer smaller than $\hbar/\tau_\varphi$ (corresponding to
the phase breaking
rate itself) do not give rise to the decoherence. Therefore, the
dephasing rate can be found from the following {\it
self-consistent} equation~\cite{AAK82}
\begin{equation}
 \frac{1}{\tau_\varphi}=\frac{k_BT}{\hbar\, g}
 \ln\frac{k_BT\tau_\varphi}{\hbar}.
 \label{eq11}
\end{equation}
The solution of this equation is shown in Fig.~\ref{fig1} by the
solid line. The product $T\tau_\varphi$ saturates with decreasing
$g$ and monotonically increases with increasing the conductance.
For the illustration purpose, in this figure we have also
presented a formal solution of Eq.~(\ref{eq11}) at $g < 1$ (dashed
line), where the equation for the dephasing rate is no longer
justified. The behavior (and even the meaning) of $\tau_\varphi$
for $g<1$ is a subtle issue and depends on the problem considered.
In principle, when the actual conductance is not very high, the
two equations, one for the conductivity and another for the
phase-breaking time, are coupled and should be solved
simultaneously.

In practice, one obtains the value of $\tau_\varphi$ from
Eq.~(\ref{eq11}) using the iteration procedure. By iteration,
starting with $\tau_\varphi^{(0)}=g\,\hbar/( k_BT ),$ one obtains
\begin{equation}
 \frac{1}{\tau_\varphi^{(1)}}=\frac{k_BT}{\hbar\, g}
 \ln g\ , \label{eq109}
\end{equation}
and so on. Usually (for $g\gg 1$) one supposes that this iteration
is sufficient for the quantitative description of the conductivity
corrections. However, as seen from Fig.~\ref{fig1} it gives fully
incorrect behavior of $\tau_\varphi$ below $g \simeq 5$, where one
expects the dephasing time to approach the value $\sim 1/T$ at
$g\sim 1$ for the case of Coulomb interaction. In the above
consideration the value of the phase breaking time depends only on
the conductivity and does not depend on other material parameters.
In this sense, $\tau_\varphi$ shows the universal behavior.

Recently, the dephasing time has been theoretically studied at
arbitrary relation between temperature and elastic mean free time
and taking into account the Fermi-liquid renormalization of the
triplet channel of Coulomb interaction~\cite{Aleiner379}. It has
been shown that in the diffusive regime ($k_B T \tau/\hbar \ll 1$)
the equation for $\tau_\varphi$ is analogous to Eq.~(\ref{eq11}):
\begin{equation}
 \frac{1}{\tau_\varphi}=
\left[1+\frac{3(F_0^\sigma)^2}{(1+F_0^\sigma)(2+F_0^\sigma)} \right]
 \frac{k_BT}{\hbar\, g}
 \ln\frac{k_BT\tau_\varphi}{\hbar}.
 \label{eq13}
\end{equation}
The only difference in this equation is a factor on the right-hand
side, which depends on the Fermi liquid constant $F_0^\sigma$. The
value of $F_0^\sigma$ can be experimentally obtained from measuring
 the logarithmic (Altshuler-Aronov) quantum correction to the conductivity
$\delta\sigma^{ee},$ caused by the e-e interaction
\cite{Altshuler, Finkelstein, ZNA},
\begin{equation}
\frac{\delta\sigma^{ee}}{G_0}=
\left[1+3\left(1-\frac{\ln(1+F_0^\sigma)}{F_0^\sigma}\right)\right]
\ln\frac{k_BT\tau}{\hbar}=K_{ee}\ln\frac{k_BT\tau}{\hbar}.\label{eqEE}
\end{equation}
In semiconductor structures, the value of $F_0^\sigma$ typically
lies within the range from $-0.5$ to $0$ (for discussion see e.g.
Ref.~\cite{sav,sav1}). For the samples investigated here,
$F_0^\sigma=-0.45...-0.25$, depending on the electron density
\cite{ourKee}. To show the difference between Eq.~(\ref{eq11}) and
Eq.~(\ref{eq13}) we have plotted the dependences
$\tau_\varphi(\sigma)$ for several $F_0^\sigma$ values in
Fig.~\ref{fig1}. It is seen that the difference increases with
conductivity increase, but even for $g=100$ it does not exceed
30~\%.

\subsection{Negative magnetoresistance and dephasing time in magnetic
field} \label{ssec:2}

How can the dephasing time be obtained experimentally? As a rule,
the value of $\tau_\varphi$ (or the ratio $\tau/\tau_\varphi$
referred further as $\gamma$) is extracted from an analysis of the
negative magnetoresistance arising due to the suppression of the
WL by a transverse magnetic field. Practically in all the cases
the experimental $\Delta\sigma(B)$-vs-$B$ curves are fitted to the
well-known expression~\cite{Hik,Schm} for the WL-magnetoconductivity (WLMC-expression):
\begin{subequations}
\label{eq20}
\begin{eqnarray}
\frac{\Delta\sigma(B)}{G_0}&=&\alpha  \left\{
\psi\left(\frac{1}{2}+\frac{1}{\tau_\varphi}\frac{\hbar}{4DeB}\right)
- \psi\left(\frac{1}{2}+\frac{1}{\tau}\frac{\hbar}{4DeB}\right)-
\ln\left(\frac{\tau}{\tau_\varphi} \right) \right\} \label{eq20a} \\[0.2cm]
&\equiv&\alpha {\cal H}(b,\gamma). \label{eq20b}
\end{eqnarray}
\end{subequations}
Here
\begin{equation}
\Delta\sigma(B)=1/\rho_{xx}(B)-1/\rho_{xx}(0),
\label{sigma-rho}
\end{equation}
$\psi(x)$ is
digamma function, $b=B/B_{tr}$, where
$B_{tr}=\hbar/(2el^2)=\hbar/(2ev_F^2\tau^2)$, and $D=v_F^2\tau/2$.
In what follows, we will consistently use the notations
$\Delta\sigma$ for the magnetoconductivity and $\delta\sigma$ for
conductivity corrections.

In the diffusive with respect to the magnetic field regime, $4 D e
B/\hbar \equiv \Omega_B \ll 1/\tau,$ one can use the asymptotics
of the second digamma function, $\psi(1/2+1/\Omega_B\tau) \sim
-\ln(\Omega_B\tau)$. Then the MR Eq.~(\ref{eq20}) can be rewritten
as a function of a single parameter $\Omega_B\tau_\varphi,$
\begin{equation}
\frac{\Delta\sigma(B)}{G_0}=\alpha  \left\{
\psi\left(\frac{1}{2}+\frac{1}{\Omega_B\tau_\varphi}\right)
+\ln\left(\Omega_B \tau_\varphi \right) \right\} \equiv \alpha
Y(\Omega_B\tau_\varphi) \label{eq20c}
\end{equation}
with the following asymptotics~\cite{Altshuler,Aleiner379}:
\begin{eqnarray}
Y(x)&=&{x^2\over 24}, \qquad\qquad \qquad\quad\quad x\to 0, \label{weakBWLMR}\\
Y(x)&=&\ln\ x+\psi(1/2)+
{\pi^2\over 2 x},
\quad  1\ll x \ll 1/\gamma,
\label{strBWLMR}
\end{eqnarray}
where $\psi(1/2)=-2\ln 2 - {\bf C}$ and  ${\bf C}=0.5772..$ is the
Euler constant. Also, using Eq.~(\ref{eq20a}) one can see that
$\Delta\sigma(B)$ saturates at $b\gtrsim 1$. The precise way of
saturation of $\Delta\sigma(B)$ depends on the character of the
disorder. In principle, the value of the dephasing time can be
obtained from the curvature of the parabolic MC in the limit of
vanishing magnetic field, $B\to 0,$ see Eq.~(\ref{weakBWLMR}).
However, usually the whole MC curve is fitted by the WLMC-formula
Eq.~(\ref{eq20}) in the range of magnetic fields where the MC is
logarithmic-in-$B$ and hence we will mainly consider the MC at
$\Omega_B\tau_\varphi>1$ in this paper.

It is worth mentioning, that the WLMC-formula Eq.~(\ref{eq20}) was
derived under the assumption that the magnetic field is
classically weak and thus does not lead to a strong
Drude-Boltzmann magnetoconductance caused by the bending of the
cyclotron trajectories. This is justified by the condition
$\omega_c\tau\ll 1,$ where $\omega_c$ is the cyclotron frequency.
For high conductances, $g\gg 1$, the logarithmic
interference-induced MC is already destroyed at much weaker
magnetic fields, $b\sim 1$, which corresponds to $\omega_c\tau\sim
1/g\ll 1$. Therefore, for $g\gg 1$ one can use the relation
(\ref{sigma-rho}). However, when the conductance is not too high,
$g\sim 1$, which is the case addressed below, the two conditions
$b=1$ and $\omega_c\tau \sim 1,$ coincide. Then the bending of
particles' trajectories may become noticeable already in the
WL-range of magnetic fields. We recall, however, that the bending
of trajectories does not give rise to the {\textit
magnetoresistance}, while the destruction of the interference
does. This is related to the fact that the interference correction
stems from the ($B$-dependent) correction to the impurity
scattering cross-section~\cite{foot-time,dmit,Groshev} and hence
renormalizes the value of the elastic scattering rate, $1/\tau$.
This is nothing but the renormalization of the longitudinal
resistivity, so that the MR arises due to the $B$-dependence of
the effective transport scattering time. This also explains why WL
effects do not give rise to the correction to the Hall
resistivity, $\rho_{xy}$: the Drude-Boltzmann expression for
$\rho_{xy}$ merely does not contain $\tau$. In other words,
Eq.~(\ref{eq20}) is in fact the correction to the
MR~\cite{Groshev} and as such is actually applicable directly to
the MR curves obtained in the experiment (without inverting the
resistivity tensor), even when the classical effect of the
magnetic field becomes visible for $g\sim 1$ at $b\lesssim 1$.

Although the prefactor
$\alpha$ has to be equal to unity within the framework of the
conventional weak-localization theory, it is always used by
experimentalists as the second fitting parameter together with
$\tau_\varphi$. An important point is that almost all
experimental data are better fitted with $\alpha < 1$,
contradicting the theory. In order to feel certain of that one
obtains the true value of $\tau_\phi$ in such a situation, it is
necessary to understand the reasons for the lowering of the prefactor in each
specific case. Possible sources for this discrepancy have been
discussed in the literature since the discovery of weak localization.
They are listed below with relevant comments.
\begin{enumerate}

\item {\it Interband scattering.} It can change the
value of $\alpha$ depending on the rate of interband transitions~\cite{Altshuler}.
The most frequently used systems where this effect is important are
$Si$-based structures of $n$-type conductivity, where there are
several valleys in the spectrum. This mechanism is not active in
our case. We will address the $n$-InGaAs quantum wells with
simplest single-valley spectrum and only one subband of the size
quantization occupied.

\item {\it Effect of ballistic paths.} Strictly speaking,
Eq.~(\ref{eq20}) was derived within the diffusion approximation.
The contributions of short trajectories, $L\lesssim l$, are
treated incorrectly (even for weak magnetic fields, $B<B_{tr}$).
Therefore, Eq.~(\ref{eq20}) is only valid under the conditions:
$\tau/\tau_\varphi \ll 1 $ and $b\ll 1$. Beyond the diffusion
approximation the MC was analyzed in a number of papers,
Refs.~\onlinecite{zyuzin,kavabata,Schm,dyak,cassam,dmit,Groshev,our1}.
The analytical expressions obtained therein are quite cumbersome
and not easy-to-use for analysis of experimental data, while the
high-field asymptotics $\delta\sigma(B)\propto 1/\sqrt{B}$ is
reached only at very strong magnetic fields, $B\gg B_{tr}$. Note
that in many papers~\cite{kavabata,Schm,dyak} the contribution of
non-backscattering processes (important in the ballistic
limit~\cite{zyuzin,dmit}, see also Appendix~\ref{gamma-vs-sigma})
was overlooked.

The applicability of Eq.~(\ref{eq20}) (with the second digamma
function not replaced by its ``diffusive'' asymptotics) beyond the
diffusion regime has been analyzed in Ref.~\onlinecite{our1} where
it has been used to fit the results of numerical simulation
(treating the numerical results like experimental data). It has
been shown that if the range of magnetic fields where the MC is
fitted using Eq.~(\ref{eq20}) includes also strong fields
$B\gtrsim B_{tr}$ [where Eq.~(\ref{eq20}) is formally no longer
justified], the resulting value of $\alpha$ will be less than
unity. Nevertheless, the value of $\tau_\varphi$ obtained in this
way happens to be close to the true one. A situation where
ballistic contribution is relevant occurs frequently in very
high-mobility structures where $B_{tr}$ is very low and can be as
small as $10^{-3}..10^{-4}$ Tesla. In what follows we will address
only the case of weak magnetic fields $B< B_{tr}$ and low
temperatures, $\tau_\varphi\gg \tau$.

\item {\it Spin relaxation.} In quantum wells with inversion
asymmetry, the Rashba or/and Dresselhaus mechanisms of spin-orbit
splitting of the energy spectrum lead to spin relaxation which
suppresses the interference--induced negative magnetoresistance in
very low magnetic fields and results in a positive MR. If this
effect is not so strong to induce the positive MR
($\tau_{\rm so}\gg \tau_\varphi$ where $\tau_{\rm so}$ is the spin-orbit
relaxation time), it can nevertheless distort the shape of MR
curve in vicinity of $B=0$ and, thus, change the parameter
$\alpha$ if the data are treated with the help of
Eq.~(\ref{eq20}). Our analysis shows that the parameters of the
best fit are unstable in this case. In particular, the value of
the prefactor strongly depends on the range of magnetic field, in
which the fit is carried out, and it is always greater than unity.
This implies that one has to exercise caution, when fitting the MC
by Eq.~(\ref{eq20a}) if even a weak spin-orbit interaction is
present in the system.
The role of spin effects in the WL was considered
for the first time in
Ref.~\onlinecite{Hik}. Using a generalized
Hikami-Larkin-Nagaoka formula~\cite{Hik}, including the spin
effects, one should obtain the value of the prefactor
as given in Ref.~\onlinecite{Hik}.
Effects of spin-orbit interaction on the WL (which are especially important
in hole systems) were further considered
in more recent papers, both theoretically and experimentally (see e.g.
Refs.~\onlinecite{pikus,Malshukov,knapandco,zduniak,GDK,Golub,Studen}
and references therein).

\item {\it Magnetic field impact on the dephasing.}  The
expression Eq.~(\ref{eq20}) was derived under the assumption that
the dephasing rate does not depend on magnetic field. As shown in
Refs.~\onlinecite{aleiner,Eiler,AAA85,Aleiner379}, the magnetic
field (rendering the inelastic processes with low energy transfer
to be inefficient) leads effectively to a decrease of the
dephasing rate. To our knowledge this effect is always ignored in
experimental papers. In Ref.~\onlinecite{tauphiB} we have analyzed
this effect both analytically and numerically. The
magnetoconductance can be described by Eq.~(\ref{eq20}) with a certain
$B$-dependent phase-breaking time $\tau_\varphi(B)$ in the first
digamma function~\cite{Aleiner379} in the whole range of magnetic
fields $b\lesssim 1$ (including the crossover region $\Omega_B\sim
1/\tau_\varphi$, not addressed accurately in
Ref.~\onlinecite{Aleiner379}). Effect of the magnetic field on the
phase breaking rate makes the negative magnetoresistance smoother
in shape and lower in magnitude than that found with the constant
phase breaking rate. Nevertheless our analysis~\cite{tauphiB}
shows that the $\Delta\sigma$-versus-$B$ plot can be well fitted
by the standard expression Eq.~(\ref{eq20}) with $\alpha\neq 1$
and a constant $\tau_\varphi$. The fitting procedure gives the
value of $\tau/\tau_{\varphi}$ which is close to the value of
$\tau/\tau_{\varphi}(B=0)$ with an accuracy of $25$\% or better
when $k_F l \gtrsim 3$ and the temperature varies within the range
from $0.4$ to $10$~K, for electron concentrations considered in
this paper.

\item {\it Electron-electron interaction in the Cooper channel.}
In low magnetic field the two interaction-induced terms can
contribute to the
magnetoresistance~\cite{Altshuler,AAKL-MIR,altshulerMT}. The first
one, known as the Maki-Thompson correction~\cite{Larkin80,MakiT},
has at $\Omega_B\ll T$ just the same $B$-dependence as the
expression Eq.~(\ref{eq20c}) but with the negative prefactor. Its
value depends on the absolute value of the effective constant of
interaction in the Cooper channel, $\lambda_c(T)$. The second term
is related to the correction to the density of states (DoS) due to
the interaction in the Cooper channel~\cite{altshulerMT} and can
be positive or negative depending on the sign of $\lambda_c(T)$,
which depends, in its turn, on the sign of the effective
interaction between electrons. The DoS-correction becomes
important at stronger magnetic fields $\Omega_B\gg T$, where it
overcomes the Maki-Thompson correction. The role of this
interaction in our experimental situation will be considered in
Sections~\ref{sec:21} and \ref{ssec:exp2}.
It will be shown that it is not the
effect of interaction in the Cooper channel that determines the
strong decrease of the prefactor $\alpha$ in the heterostructures
investigated at not very high conductance.

\item {\it Corrections of higher orders in $1/g$.} The formula
Eq.~(\ref{eq20}) is the first-order in $1/g$ correction to the
conductivity and therefore is valid only for large conductances.
Of course, there are corrections of higher orders in $1/g$ which
become important with the increase of the disorder strength or
with decreasing electron concentration. We analyze the
higher-order terms, both in the WL contribution and in the
correction induced by the mutual effect of WL and the Coulomb
interaction~\cite{aleiner}, in Section~\ref{ssec:31} and
Section~\ref{ssec:33}. This consideration allows us to find the
${\cal O}(1/g)$-corrections to the prefactor $\alpha$ and to
understand also the relation between the experimentally extracted
value of $\gamma$ and the true phase-breaking time $\tau_\varphi$
at $\sigma(B=0)\lesssim G_0$.

\end{enumerate}

In what follows, we will concentrate on the last two effects which
we believe are the most relevant sources of the reduction of the
prefactor in WLMC-expression, Eq.~(\ref{eq20}). We will show that
at not very high conductance, the effect of corrections of higher
orders in $1/g$ is more important than the effect of the
electron-electron interaction in the Cooper channel.

\section{Interaction corrections in the Cooper channel}
\label{sec:21}

It is commonly believed that it is the interaction correction
in the Cooper channel (mainly the Maki-Thompson correction to the
conductivity~\cite{MakiT,Larkin80}) which determines the reduction of the prefactor
in the MC. Indeed, in low magnetic fields the
two terms induced by the interaction in a Cooper channel contribute to the
magnetoconductance~\cite{Altshuler,AAKL-MIR,altshulerMT}
\begin{equation}
\Delta\sigma^{\rm C}_{ee}=\Delta\sigma^{\rm MT}+\Delta\sigma^{\rm DoS},
\label{eqEECor}
\end{equation}
where $\Delta\sigma^{\rm MT}$ is the Maki-Thompson correction
to the conductivity~\cite{Larkin80}
and $\Delta\sigma^{\rm DoS}$ arises due to the correction
to the DoS induced by the interaction in the Cooper
channel~\cite{Alt-Varlamov,foot-AL}.

At $B=0$ and high conductance $g\gg 1,$ for a repulsive
interaction these corrections read~\cite{Larkin80,Altshuler,AAKL-MIR,Alt-Varlamov}
(in what follows we set for brevity $k_B=\hbar=1$)
\begin{eqnarray}
\delta\sigma^{\rm MT}=G_0 \frac{\pi^2 \lambda_c^2(T)}{6}\ln(T \tau_\varphi)
=G_0 \frac{\pi^2}{6 \ln^2(T_c/T)}\ln(T \tau_\varphi),
\label{MTB=0}
\end{eqnarray}
and
\begin{eqnarray}
\delta\sigma^{\rm DoS}=-G_0\ln[\lambda_c(T)\ln(T_c\tau)]
=-G_0\ln\left[\frac{\ln(T_c\tau)}{\ln(T_c/T)}\right].
\label{MDoSB=0}
\end{eqnarray}
An important quantity governing the strength of the corrections
Eqs.~(\ref{eqEECor}), (\ref{MTB=0}), and (\ref{MDoSB=0})
is the effective amplitude of the interaction in the Cooper channel,
\begin{equation}
\lambda_c(T)=
\left[\frac{1}{\lambda_0}+\ln\frac{2 \eta}{\pi T}+{\bf C} \right]^{-1}
\equiv \frac{1}{\ln(T_c/T)},
\label{eqCEE}
\end{equation}
where $\lambda_0$ is the dimensionless ``bare'' interaction
constant, $\eta$ is the Fermi energy for Coulomb repulsion
($\lambda_0>0$) between electrons. (In the case of a
phonon-mediated attraction, $\lambda_0<0$, $\eta$ is given by the
Debye frequency.~\cite{varlamov-larkin}) Thus we have for the case
of the Coulomb repulsion (see also
Ref.~\onlinecite{varlamov-larkin} for the case of attraction)
\begin{equation}
T_c \simeq \frac{2E_F e^{\bf C}}{\pi}\exp(1/\lambda_0)>E_F.
\label{Tc}
\end{equation}
Similarly to the WL correction,
the above corrections stem from the interference of time-reversed paths and
therefore are affected by the magnetic field.
However, since the interaction is also involved in these corrections,
an additional parameter $\Omega_B/T$, relating the magnetic field and the
temperature, appears. This should be contrasted with the WL correction,
in which only the parameters $\Omega_B\tau_\varphi$ and $\Omega_B\tau$
play an important role.

For $\Omega_B\ll T,$ the Maki-Thompson correction to the
MC is given by~\cite{Larkin80,altshulerMT,Altshuler} (see also Appendix A)
\begin{equation}
\Delta\sigma^{\rm MT}\equiv\delta\sigma^{\rm MT}(B)-\delta\sigma^{\rm MT}(0)=
-G_0 \frac{\pi^2}{6 \ln^2(T_c/T)}Y(\Omega_B\tau_\varphi),
\label{MRMT-weak}
\end{equation}
where $Y(x)$ is just the same function [defined in
Eq.~(\ref{eq20c})] that describes the MC due to the suppression of
WL. Thus the Maki-Thompson correction gives rise to a parabolic MC
at $\Omega_B\ll 1/\tau_\varphi$ and to a logarithmic MC at
$1/\tau_\varphi\ll\Omega_B\ll T$. In the same range of magnetic
fields, $\Omega_B\ll T$, the DoS-correction yields a parabolic
MC~\cite{altshulerMT,AAKL-MIR}
\begin{equation}
\Delta\sigma^{\rm DoS}= -G_0 \lambda_c(T) \varphi_2(\Omega_B/2\pi T),
\label{MRDoS-weak}
\end{equation}
where the function $\varphi_2(x)$ is given by,~\cite{altshulerMT}
\begin{equation}
\varphi_2(x)= \int_0^\infty dt
\frac{t}{{\rm sinh}^2t}\left[1-\frac{xt}{{\rm sinh}(x t)}\right]=
\left\{ \begin{array}{ll} \zeta(3) x^2/4, & \quad x\ll 1, \\
                         \ln x, & \quad x\gg 1,
            \end{array}\right.\label{phi2}
\end{equation}
with $\zeta(x)$ [$\zeta(3)=1.202..$] the Riemann zeta-function.
Comparing Eqs.~(\ref{MRMT-weak}) and (\ref{MRDoS-weak}), we find
that for $1/\tau_\varphi\ll \Omega_B\ll 2\pi T \times {\rm
min}\{1,[\lambda_c(T)\ln(T\tau_\varphi)]^{1/2}\}\sim T$ the
logarithmic-in-$B$ Maki-Thompson correction to the MC dominates
over the DoS-correction.\cite{foot-MTBto0} The Maki-Thompson
correction has the same $B$-dependence as the interference
correction and effectively reduces the total prefactor in the
MC.\cite{Larkin80} The temperature dependence of $\lambda_c(T)$
translates into the $T$-dependence of the effective prefactor
$\alpha<1$ in Eq.~(\ref{eq20c}).

Let us consider these corrections at stronger magnetic fields.
Unfortunately, the exact crossover functions appear to be rather
cumbersome~\cite{Altshuler,AAKL-MIR} and we will restrict
ourselves to the analysis of the asymptotic forms of the
corrections at $\Omega_B\gg T$. As shown in Appendix A, the
Maki-Thompson contribution to the MC saturates in this range of
magnetic fields. On the other hand, it turns out that
Eq.~(\ref{MRDoS-weak}) works there as well, yielding a dominating
logarithmic-in-$B$ contribution~\cite{AAKL-MIR,altshulerMT}
\begin{equation}
\Delta\sigma^{\rm DoS}= -G_0 \frac{\ln(\Omega_B/2\pi T)}{\ln(T_c/T)}.
\label{MRDoS-strong}
\end{equation}
This result can be also obtained if for the calculation of
$\delta\sigma^{\rm DoS}(B)$ at $\Omega_B\gg 2\pi T,$ one simply
substitutes $\Omega_B/2\pi$ instead of $T$ in Eq.~(\ref{MDoSB=0}),
thus taking into account the $B$-dependence of the effective
coupling constant, which gives
\begin{equation}
\frac{\Delta\sigma^{\rm DoS}(B)}{G_0}=
\frac{\delta\sigma^{\rm DoS}(B)-\delta\sigma^{\rm DoS}(0)}{G_0}\simeq
-\ln\left[{\ln(T_c/T)\over \ln(2\pi T_c/\Omega_B)}\right]
\simeq -{\ln(\Omega_B/2\pi T)\over \ln(T_c/T)}.
\label{MDoSBstrong}
\end{equation}

We thus see that the interaction corrections in the Cooper channel
indeed reduce the effective prefactor $\alpha$ in the
MC, as compared to the non-interacting case:
\begin{eqnarray}
\alpha^{C}_{\rm ee}&=&1-{\pi^2 \over 6 \ln^2(T_c/T)},
\quad \Omega_B\ll T, \label{eq:alphaCMT} \\
\alpha^{C}_{\rm ee}&=&1-{1 \over \ln(T_c/T)},  \qquad \Omega_B\gg T.
\label{eq:alphaC}
\end{eqnarray}
Here only the asymptotics of the prefactor is presented (and we
write the corresponding conditions with the logarithmic accuracy).
To describe the crossover, one can use Eqs.~(\ref{MRDoS-weak}) and
(\ref{phi2}) for the DoS-correction, and Eq.~(\ref{MT-Hik}) for
the Maki-Thompson one, in the whole range of magnetic fields.

When the conductance is not very high (which is the situation of a
primary interest to us in Section~\ref{sec:exp}),
$\hbar/\tau_\varphi \to k_BT $ with decreasing $g$ (see
Fig.~\ref{fig1}), so that both corrections are quadratic in $B$
for $\Omega_B\ll T$. Therefore the reduction of the prefactor in
the nontrivial logarithmic MC occurring at $\Omega_B\gg
1/\tau_\varphi$ is determined by the DoS-correction and given by
Eq.~(\ref{eq:alphaC}) in this case.

\begin{figure}
\includegraphics[width=0.5\linewidth,clip=true]{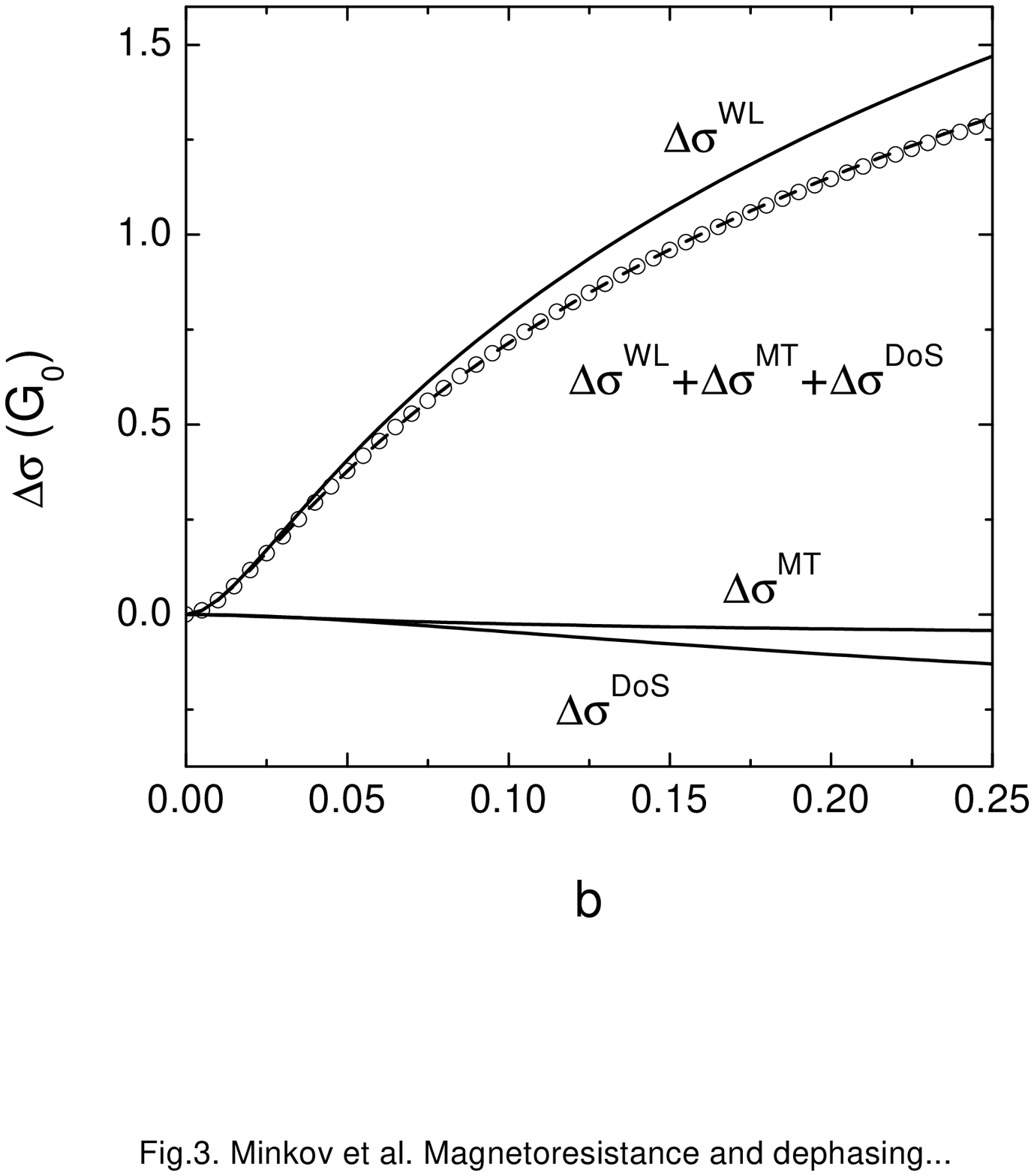}
\caption{The contributions to the magnetoconductivity in the
Cooper channel as compared with that due to the weak localization.
The $\Delta\sigma^{WL}$-versus-$b$ curve is Eq.~(\ref{eq20}) with
$\alpha=1$, $\Delta\sigma^{MT}$-versus-$b$ curve is calculated
from Eqs.~(\ref{MRDoS-weak}) and (\ref{phi2}),
$\Delta\sigma^{MT}$-versus-$b$ curve is Eq.~(\ref{MT-Hik}). The
parameters corresponding to the case of $k_F l=2.2$ (see
Section~\ref{sec:exp} for experimental details) have been used in
the calculations: $E_F= 11$~meV that corresponds to $n\simeq
2.8\times 10^{15}$~m$^{-2}$, $F_0^\sigma=-0.42$,~\cite{ourKee}
$\tau=6.2\times 10^{-14}$~s, $\gamma^{-1}=104$,  $T=1.5$~K.
Circles are the sum of all the contributions, dashed line is the
best fit by Eq.~(\ref{eq20}), which gives $\alpha=0.85$,
$\gamma^{-1}=113$. Note, the prefactor value obtained
experimentally is $0.35$ (see Fig.~\ref{fig2} and
Table~\ref{tab1}).}\label{fig:cooper}
\end{figure}

Let us now estimate the values of the prefactor $\alpha^{C}_{\rm
ee}$ corresponding to the typical parameters of our experiment.
Since we consider here the Coulomb repulsion, we have $\eta\simeq
E_F$. We also set $\lambda_0$ to be of the order of $|F_0^\sigma|$
for our estimates. All the data presented below are obtained for
electron density that changes from approximately $1\times
10^{16}$~m$^{-2}$ ($E_F\simeq 450$~K) to $2\times
10^{15}$~m$^{-2}$ ($E_F\simeq 90$~K), the value of $F_0^\sigma$
varies from $-0.25$ to $-0.45$.\cite{ourKee} Equation ~(\ref{Tc})
gives the following estimate for $T_c$: it is $3\times 10^4$~K for
the highest electron density and $10^3$~K for the lowest one.
Substituting these quantities in Eq.~(\ref{eq:alphaC}) we see that
the value of $\alpha^{C}_{\rm ee}$ is only slightly less than
unity for any electron density and temperature in the range
$(0.4-4.2)$~K: its maximal value, $\alpha^{C}_{\rm ee}\simeq
0.99$, corresponds to $n=10^{16}$~m$^{-2}$, $T=0.4$~K, and
$\Omega_B\ll 2\pi T$, the minimal value, $\alpha^{C}_{\rm
ee}\simeq 0.85$, is realized for $n=2\times 10^{15}$~m$^{-2}$,
$T=4.2$~K, and $\Omega_B\gg 2\pi T$. For reference, the
experimentally observed decrease of the prefactor is about five
times as large (see Section~\ref{sec:exp}). The contributions of
the corrections in the Cooper channel and WL-contribution in the
magnetoconductivity are illustrated by Fig.~\ref{fig:cooper}. For
calculation, we have used the parameters of one of the samples
investigated in Section~\ref{sec:exp}: $k_F l=2.2$, $E_F=11$~meV,
and $T=1.5$~K. It is clearly seen that the corrections in the
Cooper channel only slightly reduce the magnetoconductivity in
magnitude and practically does not change the curve shape. The
last is more evident if one applies the standard fitting procedure
trying to describe the total correction $\Delta\sigma^{\rm
WL}+\Delta\sigma^{\rm MT}+\Delta\sigma^{\rm DoS}$ by
Eq.~(\ref{eq20}) (compare circles and dashed line in
Fig.~\ref{fig:cooper}). This procedure demonstrates once again
that the interaction in the Cooper channel cannot be responsible
for the reduction of the prefactor in the magnetoconductivity: in
this example the interaction correction results in reduction of
$\alpha$ on the value $0.15$ instead of $0.65$ observed
experimentally.

In what follows we will consider the conductivity corrections
of higher-order in $1/g$. Taking into account such terms,
we will find the ${\cal O}(1/g)$-correction to the prefactor $\alpha$.
Therefore, in order to determine the main source of the reduction
of $\alpha$ one should compare the values of $G_0/\sigma$
and $\lambda_c(T)$. It turns out that already at sufficiently
high values of $\sigma\sim 10G_0$ the $1/g$-corrections win.
Moreover, from the theoretical point of view, the latter mechanism
of reduction of $\alpha$ will always win in the limit $T\to 0$,
since $\lambda_c(T)$ decreases while $1/g(T)$ increases with
decreasing $T$.

\section{Higher-order corrections to the magnetoconductivity}
\label{sec:3}

\subsection{Second-loop correction to the magnetoconductivity: interference term}
\label{ssec:31}
\vspace{0.1cm}

At intermediate and small values of $k_F l$ the higher order
corrections in $1/g$ should be taken into account. To sum up these
corrections, a self-consistent theory of Anderson localization was
invented in Ref.~\onlinecite{Woelfle}. The generalization of this
approach onto the case of finite magnetic field was developed in
Ref.~\onlinecite{Bryksin} (see also earlier works,
Refs.~\onlinecite{Ioshioka,Ting}). However, as will be seen below,
there is no agreement between the theory \cite{Bryksin} and
experimental results at low conductivity. This is related to the
fact that the self-consistent theory \cite{Bryksin} mistreats the
quantum corrections involving diffusons, as was pointed out in
Ref.~\onlinecite{Vollhardt-Woelfle}.

Another approach is based on the systematic analysis of
higher-order quantum corrections arising from the second-loop term
in scaling theory of
localization.\cite{GorkovLarkinKhmelnitskii,gang4} Physically, the
second-loop corrections correspond to the contributions of the
interfering waves traversing along the paths that form two loops
(instead of a single loop for the first-order WL correction) in
the real space. We start with the analysis of the next order
correction for non-interacting
electrons.\cite{GorkovLarkinKhmelnitskii} It is well
known~\cite{LeeRam,Hik81,wegner,brezin,Efetov} that the
$\beta$-function,
\begin{equation}
\beta({\mathsf g})={\partial \ln {\mathsf g} \over \partial \ln L},
\label{beta-fun}
\end{equation}
governing the scaling of the conductance with the
system size $L,$ depends on whether the magnetic
field is present (unitary ensemble)
or absent (orthogonal ensemble).
Note that in this section, {we measure the conductance per spin
in units of $G_0$}, which allows us to avoid the appearance of
additional factors of $\pi$; we will use a notation ${\mathsf g}\equiv \pi g/2$
for the such defined dimensionless conductance.
At large conductance,
only the first non-vanishing order of
the expansion of $\beta({\mathsf g})$ in powers of $1/{\mathsf g}$
is relevant.
Of course, the renormalization group equation perturbative in $1/{\mathsf g}$
can not be applied to the region of ${\mathsf g}\lesssim 1$.

Let us discuss the crossover between the orthogonal and unitary
ensembles.
In the unitary ensemble the one-loop (Cooperon)
term in beta-function vanishes, and the $\beta$-function is
given for ${\mathsf g}\gg 1$ by~\cite{wegner,Hik81,brezin}
\begin{equation}
\beta_U({\mathsf g})=-{1\over 2 {\mathsf g}^2}+{\cal{O}}\left({1\over {\mathsf g}^4}\right).
\label{betaU}
\end{equation}
Solving the scaling equation Eq.~(\ref{beta-fun}) with
Eq.~(\ref{betaU}), one gets for $L\ll \xi_U,$ i.e., for ${\mathsf
g}\gg 1$.
\begin{equation}
{\mathsf g}={\mathsf g}_0-{1\over 2 {\mathsf g}_0}\ln(L/l),
\label{unit-g}
\end{equation}
where ${\mathsf g}_0=\pi k_F l/2$.
The conductivity (in the spin-degenerate system)
is then given by
\begin{eqnarray}
\sigma=2 G_0 {\mathsf g} = \sigma_0 - {e^2\over
2\pi^2\hbar}{1\over \pi k_F l} \ln\left(\tau^D_\varphi/\tau
\right). \label{sigma2-strongB}
\end{eqnarray}
Here we use the phase-breaking length as the cutoff for the
renormalized conductance, $L=L^D_\varphi=(D\tau^D_\varphi)^{1/2}$.

In the perturbation theory, the corresponding second-loop
correction to the conductivity
\begin{equation}
\delta\sigma_{2}^{D} =
-{G_0^2\over \sigma_0}\ln\left({\tau^D_\varphi\over \tau}\right)
\label{dsigma2D}
\end{equation}
is produced by the diagrams with two and three
diffusons.\cite{GorkovLarkinKhmelnitskii} Note that the
phase-breaking time determining the $T$-dependence of the
second-order corrections $\delta\sigma_2$ is given by the
same~\cite{polyakov-samokhin} equation, Eq.~(\ref{eq11}), as
obtained for the conventional first-order WL correction
\begin{equation}
\tau^D_\varphi = \tau_\varphi.
\end{equation}

In the orthogonal ensemble
the weak-localization expression for the beta-function has the
form~\cite{wegner,Hik81}
\begin{equation}
\beta_O({\mathsf g})=-{1\over {\mathsf g}}+{\cal{O}}\left({1\over {\mathsf g}^4}\right).
\label{betaO}
\end{equation}
The term ${\cal{O}}(1/{\mathsf g}^2)$ vanishes, as
the contribution of diagrams involving Cooperons
$\delta\sigma_{2}^{C} = (G_0^2/\sigma)\ln(\tau_\varphi/\tau)$
exactly cancels the purely diffuson
(determining the result in the unitary ensemble)
contribution $\delta\sigma_{2}^{D}$.

With increasing magnetic field, the Cooperons get suppressed and
only the diffuson contribution $\delta\sigma_{2}^{D}$ survives at
$B\gg B_{tr}$, yielding the result for the unitary ensemble
discussed above. In the crossover regime between orthogonal and
unitary ensemble ($1/\tau_\varphi \lesssim B \lesssim B_{tr}$),
the positive second order Cooperon contribution can be written
similarly to the usual WL-correction:~\cite{NovGroGor}
\begin{equation}
\frac{\delta\sigma^{C}_{2}(B)}{G_0} = -{G_0\over \sigma_0} \left[
\psi\left(\frac{1}{2}+\frac{1}{\Omega_B\tau_\varphi}\right)
- \psi\left(\frac{1}{2}+\frac{1}{\Omega_B\tau}\right)
\right]. \label{coop-2}
\end{equation}
Physically, this is because the nature of the suppression of both
corrections is the same: magnetic field destroys the phase coherence between the
paths traversed in opposite directions.
Clearly, such a form matches the limiting cases $B=0$ and $B\gg B_{tr}$,
that are $\delta\sigma^{C}_{2}(0)=-\delta\sigma^{D}_{2}(0)$ and
$\delta\sigma^{C}_{2}(B\gtrsim B_{tr})\to 0,$ respectively.
Since $\delta\sigma_2^{D}$ is $B$-independent,
we have
\begin{equation}
\Delta\sigma^{WL}_{2}(B)=\delta\sigma_2(B)-\delta\sigma_2(0)=
\delta\sigma^{C}_{2}(B)-\delta\sigma^{C}_2(0)
\end{equation}
and hence the second-loop WL correction to the MC
reads
\begin{equation}
\frac{\Delta\sigma^{WL}_{2}(B)}{G_0}
=-{G_0\over \sigma_0} \; {\cal H}(b,\gamma) \label{g2coop}
\end{equation}
This expression implies that the effective prefactor
$\alpha_{\rm WL}$ depends
on the value of $\sigma_0/G_0$ (note that this is in contrast to the case of the
interaction correction in the Cooper channel,
where the prefactor is $T$-dependent),
\begin{equation}
\alpha_{\rm WL} = 1-{G_0\over\sigma_0}
\label{alpha2WL}
\end{equation}
when the two-loop interference correction is taken into account.
This is a perturbative in $1/{\mathsf g}_0$ result. In
appendix~\ref{ssec:32} we generalize this result using the scaling
approach, which would allow us to replace effectively $\sigma_0\to
\sigma$ in Eq.~(\ref{alpha2WL}) in a broad range of the
conductivity, see Section~\ref{ssec:pref}.

However, this is not the end of the story.
There also exists a two-loop correction that describes an
interplay between the weak localization and the interaction
effects. This correction is addressed in the next subsection.

\subsection{Second-loop correction to magnetoconductivity:
interplay of weak localization and interaction.}
\label{ssec:33}

Let us remind the reader, that to the leading order in $1/g$,
there are two distinct conductivity corrections. These are (i) the
WL correction, which does not involve the interaction (we assume
here that $\Omega_B^{-1}\ll \tau_\varphi,$ so that the WL
correction is cut off by the magnetic field) and (ii)
interaction-induced Altshuler-Aronov correction which is
insensitive to the magnetic field in the whole range of $B$. Both
effects give rise to the logarithmic terms in the conductivity,
\begin{equation}
\delta\sigma_{\rm WL}=G_0\ln(\Omega_B\tau), \qquad
\delta\sigma^{ee}=G_0\ln(T\tau).
\end{equation}
Note that the prefactors in front of logarithms are the same for
both corrections (for simplicity, we neglect the contribution of
the triplet channel governed by $F_0^\sigma$ in
$\delta\sigma^{ee}$, assuming that the Coulomb interaction is
weak). As a mutual effect of the interaction and weak
localization, in the next order in $1/{\mathsf g}$ there should
arise an interaction-induced and magnetic field dependent term,
$\delta\sigma^{\rm{I}\times {\rm WL}}_2$, which would also affect
the MC. For high enough temperatures, $T\gg \Omega_B,$ this
correction was calculated in Ref.~\onlinecite{aleiner}.

One can distinguish the two types of the interplay effects that
produce such a correction. The first one is the effect
of interaction-induced inelastic scattering on the WL correction.
The corresponding correction is termed $\delta\sigma_{\rm deph}$
in Ref.~\onlinecite{aleiner} and is related to the $B$-dependent dephasing time.
The second effect can be thought of as
the influence of weak localization
on the interaction-induced Altshuler-Aronov correction,
the corresponding correction being termed
$\delta\sigma_{\rm CWL}$.

In what follows, we will analyze the
interaction correction to the MC at
$\Omega_B\tau_\varphi\gg 1$.
For $1/\tau_\varphi \ll \Omega_B\ll T$,
the dephasing term is given by~\cite{aleiner}
\begin{equation}
\delta\sigma_{\rm deph}=\frac{G_0^2}{\sigma_0} \left\{\frac{\pi T
}{\Omega_B}\left[\ln\frac{T }{\Omega_B} +1 \right]
+\ln\frac{1}{\Omega_B\tau}
 \right\}, \qquad T \gg \Omega_B,
 \label{eq22a}
\end{equation}
while the cross-term
of Coulomb
interaction and weak localization looks as follows \cite{aleiner}
\begin{equation}
\delta\sigma_{\rm CWL}=\frac{G_0^2}{\sigma_0}\left\{
 \frac{1}{2}\ln\left(\frac{1}{\Omega_B\tau}\right) + {\cal O}(\ln T\tau)
 \right\}, \qquad T\gg \Omega_B.
 \label{eq22aa}
\end{equation}
The term ${\cal O}(\ln T\tau)$ is beyond the accuracy of the theory,
since the second-order interaction correction (not involving Cooperons)
produces an analogous contribution. This term, however,
does not depend on the magnetic field and
we throw it away when the MR is considered.

We see that in the range of high enough temperature, $T \gg
\Omega_B$, apart from the modification of the dephasing rate by
the magnetic field,\cite{footnote} described by the first term in
Eq.~(\ref{eq22a}), there is a logarithmic contribution to the MC,
\begin{equation}
\delta\sigma^{\rm{I}\times {\rm WL}}_2=\frac{3 G_0^2}{2 \sigma_0}
 \ln\left(\frac{B_{tr}}{B}\right), \qquad T \gg \Omega_B.
 \label{eq22b}
\end{equation}
This contribution is very similar to that found in the preceding
subsections and also reduces the prefactor $\alpha$ in front of
the logarithmic term. However, in this range of magnetic fields
the first term in Eq.~(\ref{eq22aa}) corresponding to the
$B$-dependent~\cite{aleiner,Aleiner379} dephasing time
(Sec.~\ref{ssec:2})
\begin{equation}
{1\over \tau_\varphi(B)}\simeq {T\over g}\ln(T/\Omega_B),
\qquad 1/\tau_\varphi\ll \Omega_B \ll T,
\label{tfb}
\end{equation}
dominates and the
subleading logarithmic term Eq.~(\ref{eq22b}) as well as the
second-loop WL-contribution are of little importance. We will
analyze the role of the contribution Eq.~(\ref{eq22a}) in more
detail elsewhere.\cite{tauphiB} In experiments discussed in
Sections~\ref{sec:exp} and \ref{sec:exp1}, the fitting of the MC
is carried out in the range of magnetic field such that $T\ll
\Omega_B$ and therefore the magnetic-field impact on the dephasing
is of a less importance in our case. Note also that with
decreasing ${\mathsf g}$ the above range $1/\tau_\varphi \ll
\Omega_B\ll T$ tends to shrink.

In stronger magnetic fields (or, equivalently, of lower
temperatures $T\ll \Omega_B$, not considered in
Ref.~\onlinecite{aleiner}), the situation changes in the following
way:\cite{gornyi-unpub} the magnetic-field dependent contribution
to the dephasing term becomes small, $\propto (T/\Omega_B)^2$,
since the corresponding frequency integral is determined by
$\omega\lesssim T \ll \Omega_B$. Therefore, the main contribution
to the MC comes from $\delta\sigma_{\rm CWL}$. This contribution
reads~\cite{gornyi-unpub}
\begin{equation}
\delta\sigma_{\rm CWL}=\frac{G_0^2}{\sigma_0}
\ln\left(\frac{1}{\Omega_B\tau}\right), \qquad T\ll \Omega_B,
 \label{eqCWL}
\end{equation}
and, therefore,
\begin{equation}
\delta\sigma^{\rm{I}\times {\rm WL}}_2\simeq
\delta\sigma_{\rm CWL}=\frac{G_0^2}{\sigma_0}
 \ln\left(\frac{B_{tr}}{B}\right).
 \label{eqIWL-strong}
 \end{equation}
Similarly to the one-loop corrections, the interaction-related
contribution Eq.~(\ref{eqIWL-strong}) and ``noninteracting" WL
correction Eq.~(\ref{g2coop}) have the same prefactors in front of
$\ln(B_{tr}/B)$. It is worth mentioning that the logarithm-squared
terms of the types $\ln^2[1/(\Omega_B\tau)]$ and
$\ln(T\tau)\ln[1/(\Omega_B\tau)]$ do cancel out at $T\ll
\Omega_B$, as in the case of weaker magnetic field considered in
Ref.~\onlinecite{aleiner}. Note that the interaction-based
renormalization group (RG) equations derived by
Finkelstein~\cite{Finkelstein} are the one-loop equations with
respect to the disorder, while here we are dealing with the
second-loop contribution.

When the parameter $\sigma/G_0$ is large, the two-loop
correction is
much less than the absolute value of the first-order correction
$\sigma^{WL}_1$ which in the same magnetic field range is
\begin{equation}
 \frac{\delta\sigma_1^{WL}}{G_0}=-\ln\left(\frac{B_{tr}}{B}\right).
 \label{eq221}
\end{equation}
When $\sigma$ decreases, both $\delta\sigma_2^{\rm{I}\times {\rm WL}}$
and $\delta\sigma_2^{WL}$ become more important, and the resulting
conductivity correction,
$\delta\sigma=\delta\sigma_1^{WL}+\delta\sigma_2^{WL}+
\delta\sigma_2^{\rm{I}\times {\rm WL}},$ looks as follows
\begin{equation}
 \frac{\delta\sigma}{G_0}\simeq -\left[1-(1+1)\frac{G_0}{\sigma_0}\right]
 \ln\left(\frac{B_{tr}}{B}\right), \qquad T\ll \Omega_B.
 \label{eq23}
\end{equation}
Moreover, as in the case of the ``non-interacting" WL terms discussed in the
preceding subsection,
we can replace $\sigma_0\to \sigma(b=1)$ in the above equation.
Thus, in the second-loop order, the combined effect of weak-localization
and Coulomb interaction reduces the prefactor $\alpha$ in front
of the logarithmic correction to the MC:
\begin{equation}
\alpha_2=1-\frac{2 G_0}{\sigma}.
\label{alpha2}
\end{equation}
This is one of the central results of the present paper.

\subsection{Meaning of the dephasing time extracted from experiments}
\label{ssec:tauphiloc}

In the preceding subsections we have analyzed the role of the
second-loop corrections to the conductivity, $\delta\sigma_2(B)$.
It has been demonstrated that these corrections give rise to a
reduction of the effective prefactor in the WLMC-expression
Eq.~(\ref{eq20a}). On the other hand, when both the zero-$B$
[$\sigma(b=0)$] and the strong-$B$ [$\sigma(b\gg 1)$]
conductivities are still larger than $G_0,$ the second-order terms
do not affect significantly the value of the dephasing time
extracted from fitting MC by the WLMC-expression. This is, in
particular, because of the fact that the phase-breaking time
governing the $T$-dependence of second-loop conductivity
corrections is equal to the ``one-loop'' dephasing
time.\cite{polyakov-samokhin} Therefore it becomes possible to
attribute the experimentally obtained value of $\gamma_{\rm fit}$
to the true value of dephasing time,
$\tau_\varphi=\tau/\gamma_{\rm fit}$ in the range of moderately
``high'' conductivities, $\sigma(b=0)\gtrsim 3G_0$ and for all the
experimentally accessible temperatures.

Let us now discuss the relation between quantity $\gamma_{\rm fit}$
and the real dephasing time in a broader range
of $\sigma(b=0)$, including the WI regime, where $\sigma(b=0)< G_0$
whereas $\sigma(b\gg 1)>G_0$. We will demonstrate that in the WI regime
the value of $\gamma_{\rm fit}$ obtained from the fitting procedure is
not proportional to the dephasing rate.
This also may affect the experimentally obtained
value of $\tau_\varphi$ in the crossover between the WL and WI regimes.

In Appendix~\ref{gamma-vs-sigma}, we show that the fitting of the MC with the use
of the WLMC-formula gives the following value of the parameter $\gamma$:
\begin{equation}
\gamma_{\rm fit}={\cal C}
\exp\left\{{1\over \alpha G_0}\Big[\sigma(b=0)-\sigma(b\gg 1)\Big]\right\},
\label{gamma-sigma}
\end{equation}
where the numerical factor of order unity, ${\cal C},$ is related
to $T$-independent contribution of ballistic paths and thus
depends on the nature of disorder. In the case of a white-noise
disorder, ${\cal C}=1/2$.\cite{GDK}

The equation (\ref{gamma-sigma}) holds for
large $\sigma(b\gg 1)$ but for an arbitrary $\sigma(b=0).$
Here $\sigma(b=0)$ and $\sigma(b\gg 1)$ are the {\it total} conductivities, including
e.g. interaction-induced contributions.
When the conductivity is high, $\sigma(b=0)\gg G_0$, it is sufficient to
consider the one-loop corrections to the conductivity.
Then we have $\alpha\simeq 1,$
$\sigma(b\gg 1)\simeq \sigma_0+\delta\sigma^{ee},$ and
$\sigma(0)=\sigma_0+G_0\ln(2 \gamma)+\delta\sigma^{ee},$
which yields
\begin{equation}
\gamma_{\rm fit}\simeq \gamma=\tau/\tau_\varphi,\qquad g\gg 1,
\label{gammaWL}
\end{equation}
in accordance with the standard WL-theory.

Let us now consider the WI regime. In this regime, the quantum
corrections are strong and almost compensate the Drude
conductivity. Let us first consider an ideal but rather a
non-realistic situation of large $\sigma_0\gg G_0$ {\it and
exponentially low} temperatures, such that $\sigma_0/G_0\ll
\ln(\tau_\varphi/\tau)\ll(\sigma_0/G_0)^2$. In this case we can
set $\sigma(b=0)\simeq 0,$  $\alpha\simeq 1,$ and substitute
$\sigma_0$ for $\sigma(b\gg 1)$ in Eq.~(\ref{gamma-sigma}):
\begin{equation}
\gamma^{\rm (WI)}_{\rm fit}\sim
\exp\left\{-{\sigma_0\over \alpha G_0}\right\}\sim
\left({l \over \xi_O}\right)^2
\label{gammaWI}
\end{equation}
Obviously, the quantity $\gamma^{\rm (WI)}_{\rm fit}$ from
Eq.~(\ref{gammaWI}) has nothing to do with the true value of the
dephasing time. In particular, the ``experimentally obtained''
phase-breaking time, $\tau_\varphi^{\rm fit}=\tau/\gamma_{\rm
fit},$ saturates with decreasing $T$ at the value given by the
localization length: $\tau_\varphi^{\rm fit}\sim \xi^2_O/D,$
whereas the real $\tau_\varphi(T)$ diverges in the limit $T\to 0.$

This has the following simple explanation.
When $\sigma(b=0)\ll G_0$ and $\tau_\varphi\gg \xi_O^2/D,$ electrons are
localized and only for $l_B\sim \xi_O$ their motion becomes diffusive
at scales larger than $\xi_O$.
On the other hand, as mentioned in the Introduction, the magnetic field
gives rise to a parabolic MR (whatever the mechanism of the MR is)
even in the localized regime.
Thus, the parabolic low-field MR persists up to the field
for which $\Omega_B\tau_\varphi\gg 1$.
Only at stronger fields the MR becomes logarithmic.
From ``the point of view'' of the WLMC-expression, this
indeed corresponds to
$\gamma^{\rm (WI)}_{\rm fit}\sim (l/\xi_O)^2$.
This is because the fitting procedure yields the value $1/\tau_\varphi^{\rm fit},$
related to the strength of magnetic field at which
the crossover between $B^2$ to $\ln B$ behavior of the MC occurs.

In the realistic situation of intermediate conductances and not
too low $T$, the temperature behavior of $\gamma_{\rm fit}$
appears to be very complicated in the WI-regime. In particular,
even at moderately low temperatures, the experimentally extracted
value of $\gamma_{\rm fit}$ may scale with the temperature as
$T^p$ with $p \neq 1$, if the conductance is not very high.
Moreover, since the $T$-dependence of $\gamma_{\rm fit}$ in the WI
regime is mainly determined by the $T$-dependence of
$\sigma(b=0,T)$ in the localized regime, the behavior of
$\gamma_{\rm fit}(T)$ depends strongly on the concrete mechanism
of transport in the localized regime. Qualitatively, the dephasing
rate extracted from the experiment can be roughly approximated to
match Eqs.~(\ref{gammaWL}) and (\ref{gammaWI})
\begin{equation}
{1\over \tau_\varphi^{\rm fit}(T)}\sim {1\over \tau_\varphi(T)} +
{1\over \tau}\left( { l \over \xi_O }\right)^{2/\alpha}.
\label{gamma-empirical}
\end{equation}
However, this formula does not allow one to describe quantitatively
the $T$-dependence of the true dephasing time $\tau_\varphi(T)$ in the low-$T$
regime.

The relation between the real dephasing rate and the behavior of
$\gamma_{\rm fit}(T)$ can be illustrated using the following toy
model. Let us assume that the true dephasing rate is always
proportional to temperature, independently of the value of the
conductance,
\begin{equation}
\gamma_{\rm true}(T)\equiv T/T_0.
\end{equation}
Furthermore, for simplicity we consider only the interference
contribution to the conductivity, that is we neglect the
corrections discussed in Section~\ref{ssec:33}. We also neglect
the $T$-independent ballistic contributions, so that the numerical
factor in Eq.~(\ref{gamma-sigma}) is equal unity. Then the
conductivity at high magnetic fields is given by
Eq.~(\ref{sigma2-strongB}),
\begin{equation}
\sigma(b\gg 1;T)=\sigma_0-{G_0^2\over \sigma_0}\ln(T_0/T),
\label{sigmastrongB}
\end{equation}
while the prefactor in the WLMC-formula decreases
as $\alpha_{WL}=1-G_0/\sigma(b\gg 1).$
The zero-$B$ conductivity is described by
\begin{equation}
\sigma(b=0;T)=\sigma_0-G_0 \ln(T_0/T), \qquad \sigma(b=0)/G_0>1
\label{sigmab=0WL}
\end{equation}
in the WL-regime, specifically for
\begin{equation}
T > T_1\equiv T_0 \exp(\sigma_0/G_0-1).
\label{T1}
\end{equation}
In the localized regime ($T<T_1$ corresponding to $\sigma(b=0)<G_0$), we assume that
the conductivity in our toy-model is due to some activation mechanism
(which is not the case in our experiments described below, but
nevertheless reflects qualitatively the behavior of the zero-$B$ conductivity,
when it is small),
\begin{equation}
\sigma(b=0;T)=G_0\exp(1-T_1/T), \qquad \sigma(b=0)/G_0<1.
\label{sigma-activ}
\end{equation}
Remarkably, the two expressions Eq.~(\ref{sigmab=0WL}) and
Eq.~(\ref{sigma-activ})
 for $\sigma(b=0)$ match
each other very nicely and are almost indistinguishable in the
range $0.5\lesssim \sigma(b=0)/G_0 \lesssim 1.5$ for arbitrary
$\sigma_0$. We substitute these conductivities in
Eq.~(\ref{gamma-sigma}) with ${\cal C}=1$ and plot the value of
$\gamma_{\rm fit}$ which would be obtained in an experiment on our
toy-system.

\begin{figure}
\includegraphics[width=\linewidth,clip=true]{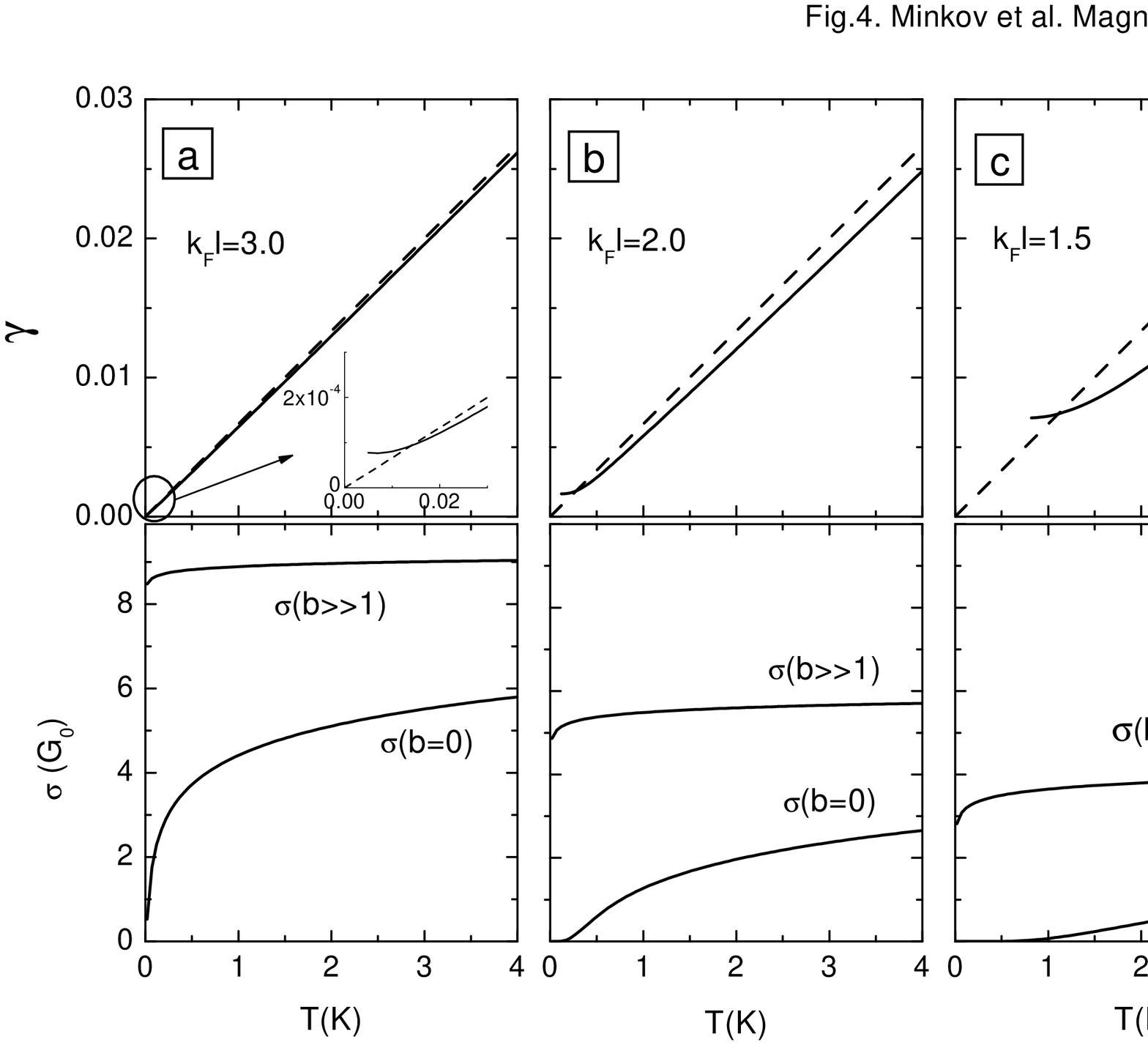}
\caption{Illustration of low-temperature saturation of
$\tau_\varphi$ obtained ``experimentally" for the toy-model with
different $k_Fl$-values: (a)~--~$k_Fl=3$ ($T_1=0.033$~K),
(b)~--~$k_Fl=2$, ($T_1=0.76$~K), and (c)~--~$k_Fl=1.5$
($T_1=3.7$~K). Upper panels show the temperature dependences of
$\gamma_{\rm true}=T/T_0$, $T_0=150$~K (dashed lines) and
$\gamma_{\rm fit}$ (solid lines) found from
Eq.~(\ref{gamma-sigma}) with the use of $T$-dependence of
$\sigma(b=0)$ and $\sigma(b\gg 1)$  shown  in lower panels. }
\label{fig:toy}
\end{figure}

The results of such an experiment are shown in Fig.~\ref{fig:toy}.
It is clearly seen that the ``experimentally extracted'' dephasing
rate deviates in the low-$T$ limit from the true one, which by
definition is described by a straight line. Moreover, the
saturation of the dephasing, occurring when $\sigma(b=0)<G_0,$
becomes evident at sufficiently low temperatures even for high
enough Drude conductivities (the higher is the conductivity, the
lower is the ``saturation temperature''). On the other hand, at
high temperatures, $\gamma_{\rm fit}$ is linear-in-$T,$ for all
$\sigma_0$, implying that the fitting procedure gives a reasonable
high-$T$ behavior of the dephasing time.

We conclude that the fitting of the MC with the use of the WLMC-expression
cannot serve to obtain the real temperature behavior of the phase-breaking time
at $\sigma(b=0)\lesssim G_0.$ Moreover, at such low conductivities the fitting procedure
may yield seemingly a spurious saturation of the ``experimentally extracted''
dephasing rate at $T\to 0$.

\subsection{Effect of second-loop corrections to the MC: summary}
\label{ssec:34}

We can  summarize the results of the preceding sections as follows:
\begin{enumerate}

\item  {\it $B=0$; WL contribution.} The terms of the second and third
orders in $1/g$ cancel out in the relative interference correction in zero
magnetic field.
This means that for numerical reasons the
temperature dependence of $\delta\sigma_{\rm WL}$ at $B=0$ is
experimentally just the same as for the case $g \gg 1$,
down to low enough values
of $\sigma \simeq (2-3)G_0$:
\begin{equation}
 {\delta\sigma_{\rm WL}(T) \over G_0}= -\beta \ln\left(\tau_\varphi(T)\over
 \tau\right), \mbox{   where   } \beta =1.
 \label{eqBeta}
\end{equation}

\item {\it $B\neq 0$; no interaction.} The terms of the second
order in $1/g$ do not influence the shape of the magnetic field
dependence of the interference correction leading only to the
decreasing of the prefactor. Therefore, the MC due to suppression
of the WL is described by Eq.~(\ref{eq20a})
\begin{equation}
{\Delta\sigma(b)\over G_0}=\alpha_{WL} \left\{ \psi\left({1\over
2}+{\gamma \over b}\right) - \psi\left({1 \over 2}+{1\over
b}\right)- \ln\gamma \right\},
 \label{eq2020}
\end{equation}
with the prefactor $\alpha_{WL}$ decreasing as $\alpha_{WL} = 1-G_0/\sigma$
with lowering
$\sigma$ down to $\sigma\simeq (2-3)G_0$.

\item {\it Coulomb interaction; $1/\tau_\varphi\ll \Omega_B\ll
T$.} The combined effect of the weak localization and Coulomb
interaction leads to magnetic-field dependent corrections to the
conductivity of the same order in $1/g$ as in previous case. At
high temperatures, $T \gg \Omega_B$, the main effect is in the
$B$-dependence of the dephasing time, which is reflected in the
correction to the high-$B$ asymptotics of the digamma-function
formula, Eqs.~(\ref{eq20})~--~(\ref{strBWLMR}),
\begin{equation}
\delta[\Delta\sigma(B)]\sim {\pi^2 \over 2} \left[{1\over
\Omega_B\tau_\varphi(B)}-{1 \over \Omega_B\tau_\varphi(0)}\right]
\simeq {\pi^2  \over 2 g} {T \over \Omega_B} \ln\left({T
\over g \Omega_B} \right). \label{tauphiH-corr}
\end{equation}
Also, in this range of $B$, the Maki-Thompson
correction to the MC dominates over the
DoS-correction in the Cooper channel.

\item {\it Coulomb interaction, stronger magnetic field, $T\ll
\Omega_B\ll 1/\tau$.} The combined effect of the weak localization
and Coulomb interaction yields a logarithmic contribution to the
MC, $\delta\sigma=(G_0/\sigma)\ln[1/(\Omega_B\tau)]$. Therefore,
the total prefactor $\alpha$ in the MC is given by
$\alpha_2=1-2G_0/\sigma$. This suggests that if the MC is fitted
by Eq.~(\ref{eq20}), in the range of magnetic fields $T\ll
\Omega_B\ll 1/\tau$, the decrease of the prefactor $\alpha$ is due
to the second-loop corrections. Note, that taking into account the
contribution of the triplet channel to $\delta\sigma_{\rm CWL}$
(neglected above) reduces the contribution of e-e interaction to
the prefactor of the logarithmic conductivity correction,
similarly to the case of the first-order Altshuler-Aronov
correction~\cite{gornyi-unpub}. We also recall that the correction
due to interaction in the Cooper channel is dominated by the DoS
correction at such magnetic fields.

\item $B>B_{tr}$. In this range of magnetic fields
the logarithmic corrections to the MC vanish.
However, there are $B$-independent corrections
$\propto (G_0/\sigma)\ln(T\tau)$
coming from the second-loop contributions,
both from the non-interacting contribution
(WL in the unitary ensemble) and from the cross-term (Coulomb plus WL).
These corrections are important at low enough conductivities,
when they can give an appreciable contribution to the prefactor
of the $T$-dependence of the high-$B$ conductivity.

\item {\it Dephasing time.}
The fitting of the MC by the WLMC-expression
gives a correct value of the dephasing rate for $\sigma(B=0)\gtrsim 3G_0$.
The $T$-dependence of $\tau_\varphi$ is given by a solution of the self-consistent
equation (\ref{eq11}) rather than by the first iteration of this equation
for intermediate conductances.
When applied in the WI regime, $\xi_O<L_\varphi<\xi_U$,
the WLMC-expression yields the value of $\gamma_{\rm fit}$ which is
not proportional to the true dephasing rate, but contains information about the
localization length, $\xi_O.$
\end{enumerate}

\begin{figure}
\includegraphics[width=0.7\linewidth,clip=true]{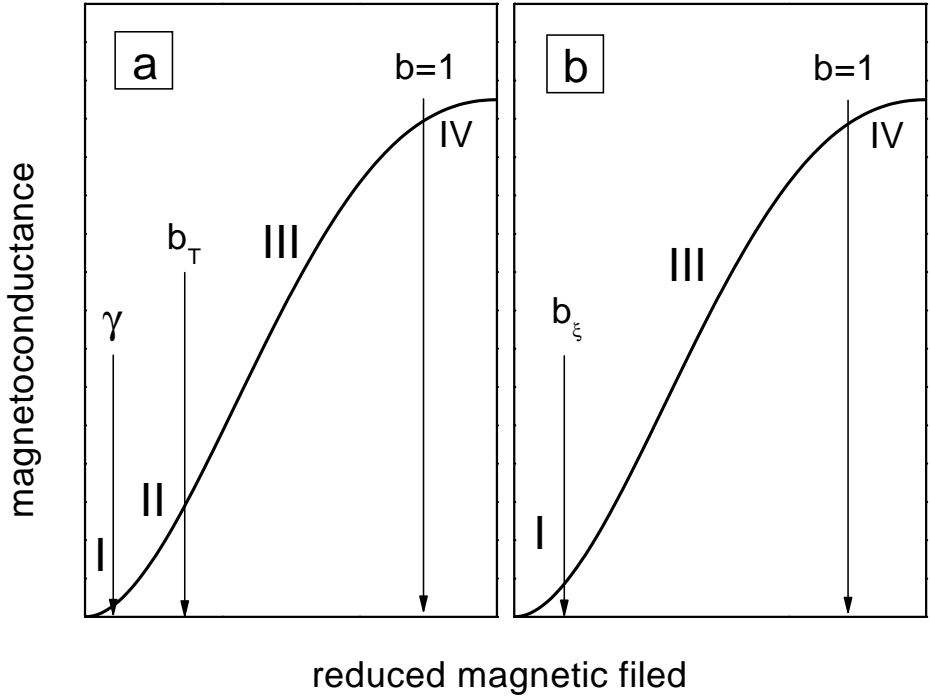}
\caption{Schematic representation of the low-field quantum
magnetoconductivity. Arrows show characteristic magnetic fields
which mark off the regions with different behavior of
magnetoconductivity (see text). It should be emphasized that the
region III is much wider than regions I and II in our case (see
Table~\ref{tab1}).} \label{fig:MFRegions}
\end{figure}

The above results are illustrated in Fig.~\ref{fig:MFRegions}.
In the WL-regime, the
magnetoconductivity as a function of $b$ behaves differently in
the four regions of the magnetic field, ${\rm I:}\ b<b_\varphi, \ {\rm
II:}\ b_\varphi<b<b_T,\ {\rm III:}\ b_T<b<1,$ and ${\rm IV:}\
b>1$. Here $b_\varphi=\tau/\tau_\varphi=\gamma$
is given by the dephasing rate and $b_T=T\tau$ is set by the temperature.
In the region I, the MC is quadratic. In the region II, the deviation
from the WLMC-formula Eq.~(\ref{eq20}) is determined by the impact
of the magnetic field on the dephasing, and therefore other
second-loop corrections are irrelevant.
The region II, however, shrinks to zero with decreasing conductance,
$\sigma\to G_0$. In the region III, the
$B$-dependence of the dephasing is no longer crucial. The MC is
given by Eq.~(\ref{eq20}) and the value of the prefactor $\alpha$
is determined by the second-loop contributions,
Eq.~(\ref{alpha2}). If the fitting procedure is carried out in the
range of $B$ involving the fields such that $b\gg b_T,$ it is this
value of the prefactor which is expected to be found
experimentally.
In the WI-regime, the region II disappears, while the value of the
magnetic field $b_\xi$, where the crossover between the parabolic (region I)
and logarithmic (region III) MC occurs, is determined by the
localization length $\xi_O$: $b_\xi\sim(l/\xi_O)^2$.
The MC in the region I ($b<b_\xi$) is beyond the scope of the present paper.
On the other hand, the MC in the region III has the same origin as in the WL-regime
and can be fitted by Eq.~(\ref{eq20}).

In what follows we present the experimental results obtained in a
wide range of conductivity. We start our analysis from the simpler
case of high conductivity and follow what happens with the
WL and MC with changing $T$ and $k_Fl,$ and thus with the
decreasing of the conductivity.
We compare the experimental results with the above theory
and find a quantitative agreement between the theory and the experiment.

\section{Experiment}
\label{sec:exp}

In order to test quantitatively such refined theoretical
predictions as presented above, suitable two-dimensional
structures have to be used. First of all, the structures should be
based on materials with single valley energy spectrum. Only one
size-quantized subband should be occupied. Electrons should be
only in the quantum well, no electrons should be in the doping
layers. Finally, to avoid spin-dependent effects, the structures
have to be symmetrical in shape in the growth direction. The
single quantum well heterostructures based on $A_3B_5$
semiconductors met  these requirements. We have investigated three
types of the GaAs/In$_x$Ga$_{1-x}$As/GaAs single quantum well
structures. They are distinguished by a  ``starting'' nominal
disorder that is achieved by a different manner of doping.

\subsection{Experimental details and samples}
\label{sec:sampl}

The heterostructures with 80\AA-In$_{0.2}$Ga$_{0.8}$As single
quantum well in GaAs were grown by metal-organic vapor-phase
epitaxy on a semi-insulator GaAs substrate. Structure H451 with
high starting disorder had $Si-\delta-$doping layer in the center
of the quantum well. The electron density $n$ and mobility $\mu$
in this structure were $n=0.89\times10^{16}$~m$^{-2}$ and
$\mu=0.23$~ m$^2$/Vs. Structure Z88 had lower starting disorder
because the doping $\delta$ layers were disposed on each side of
the quantum well and were separated from it by the 60~\AA\ spacer
of undoped GaAs. The parameters of structure Z88 were
$n=5.1\times10^{15}$~m$^{-2}$ and $\mu=1.3$~ m$^2$/Vs. Finally,
the third structure 3509 had not $\delta$ doping layers. The
conductivity of this structure was less than $10^{-2}\,G_0$ at
liquid helium temperatures. The thickness of undoped GaAs cap
layer was 3000 \AA\ for all structures. The samples were mesa
etched into standard Hall bars and then an Al gate electrode was
deposited by thermal evaporation onto the cap layer of the
structures H451 and Z88 through a mask. Varying the gate voltage
$V_g$ from $0.0$ to $-3..-4$~V we decreased the electron density
in the quantum well and changed $k_F l$ from $9-30$, for different
samples, down to $\simeq 1$ (the values of $k_F l$ and $B_{tr}$
have been experimentally  found as described in
Appendix~\ref{sec:kfl}). The conductivity of structure 3509 was
changed via illumination by light of a incandescent lamp through a
light guide. Due to persistent conductivity effect we were able to
increase the conductivity and electron density for this structure
up to approximately $60\,G_0$ and $5\times 10^{15}$~m$^{-2}$,
respectively, changing the duration and intensity of illumination.
Several samples of each structures have been measured and they all
demonstrate the universal behavior.

\subsection{Overview of the experimental results}
\label{ssec:exp}

\begin{figure}
\includegraphics[width=1\linewidth,clip=true]{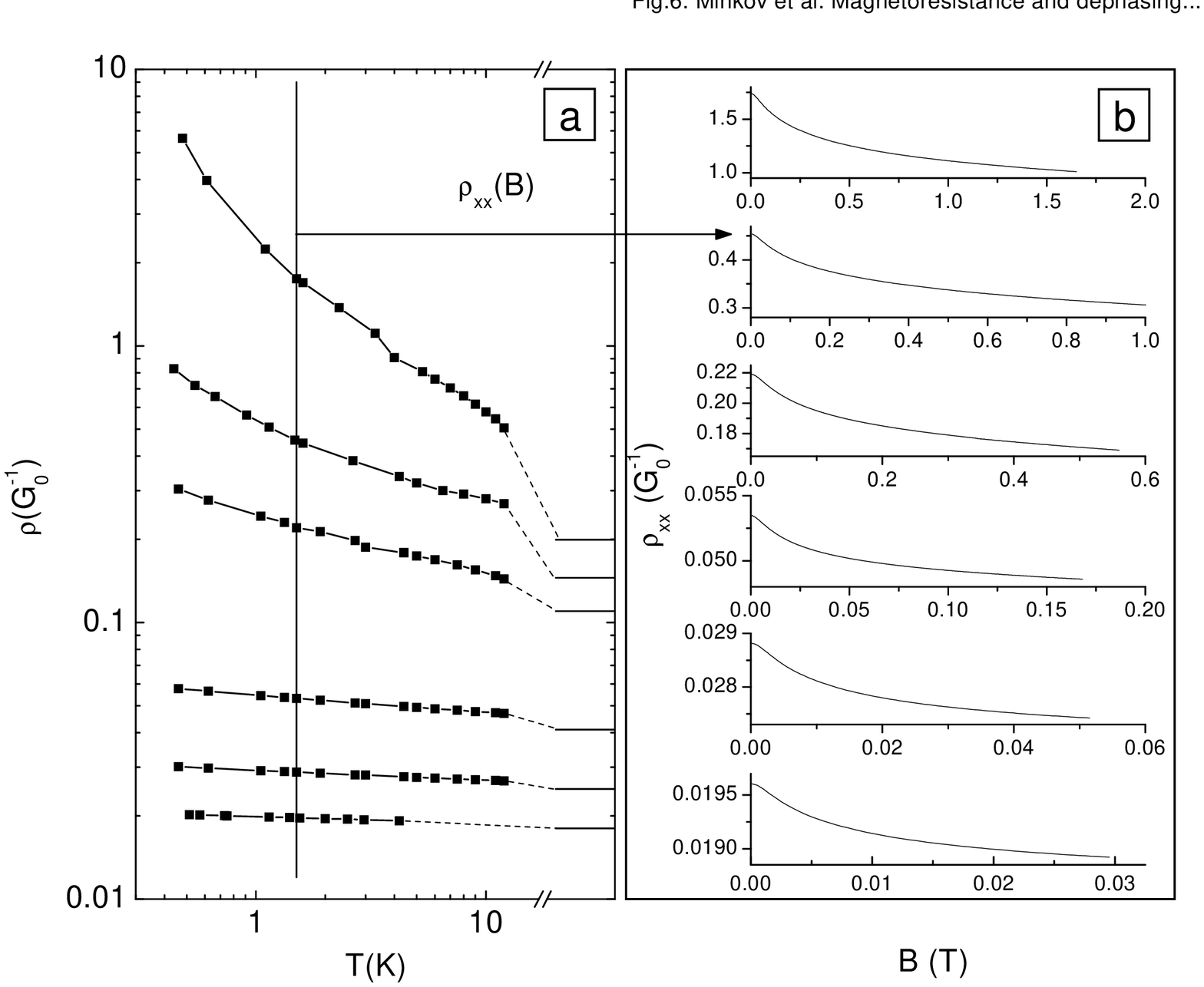}
\caption{(a) The temperature dependence of the resistivity for
structure Z88, measured at different $k_F l$ values: 1.6; 2.2;
2.9; 7.7; 12.8; 17.9 (from the top to the bottom). Horizontal
lines show the values of $\sigma_0^{-1}=(\pi k_F l G_0)^{-1}$
where $k_F l$ was found as described in Appendix \ref{sec:kfl}.
(b) The $\rho_{xx}$-versus-$B$ dependences measured for $k_F l$
from the left panel at $T=1.5$~K.} \label{rho(T)}
\end{figure}

The temperature dependences of the zero-$B$ resistivity $\rho$
measured at several $k_Fl$-values controlled by the gate voltage
for one of the samples made from structure Z88 are presented in
Fig.~\ref{rho(T)}(a). A thorough analysis of these dependences has
been done in Ref.~\onlinecite{WLtoSL}. It has been shown that the
$\sigma$-versus-$T$ dependences are close to the logarithmic ones
for $k_F l \gtrsim 2$ over the actual temperature range. For the
lower $k_Fl$- values, when the conductivity is less than $e^2/h,$
a significant deviation from the logarithmic behavior is observed.
The temperature dependences of conductivity are well described
within the framework of the conventional theory of the quantum
corrections down to $k_F l\simeq 2$.  It has been also shown that
the interference contribution to the conductivity for $B=0$
exceeds the contribution due to the electron-electron interaction
in $3-5$ times.

\begin{figure}
\includegraphics[width=0.7\linewidth,clip=true]{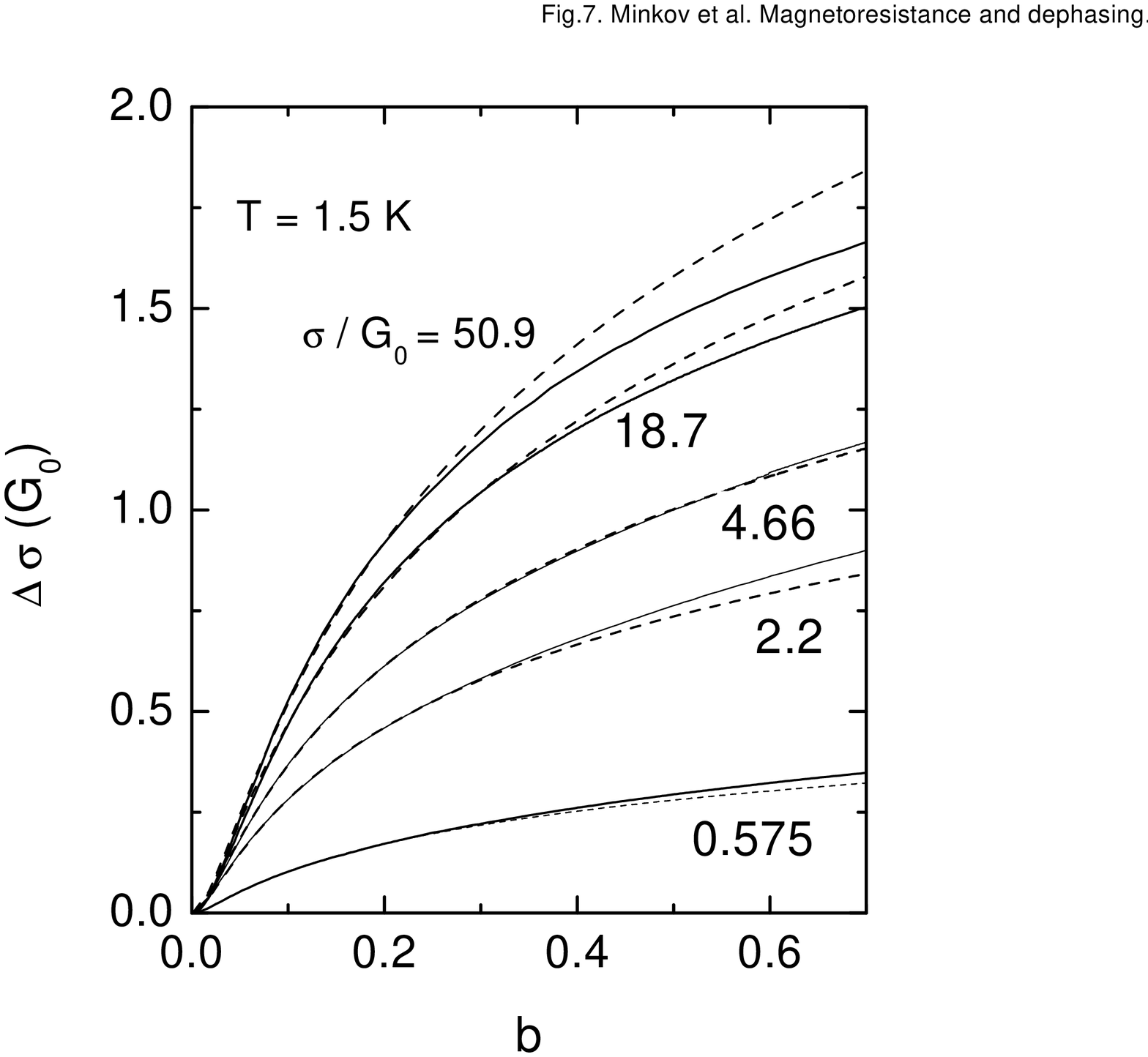}
\caption{(a) The value of $\Delta\sigma$ as a function of reduced
magnetic field, $b=B/B_{tr}$, measured for structure Z88 at
$T=1.5$~K and different $k_F l$ values. Solid curves are the
experimental data, dashed curves are the best fit by
Eq.~(\ref{eq20}) with $\gamma$ and $\alpha$ given in
Table~\ref{tab1}. } \label{fig2}
\end{figure}

The experimental magnetic-field dependences of $\rho_{xx}$
measured at $T=1.5$~K for the different $k_F l$-values are
presented in Fig.~\ref{rho(T)}(b). We restrict our consideration
to the range of low magnetic field. In these fields the negative
MR is completely determined by the interference effects which is
subject of this paper. The high-magnetic-field MR and the role of
electron-electron interaction have been studied in details in
Ref.~\onlinecite{ourKee} and we will not consider them here. Even
a cursory examination of Fig.~\ref{rho(T)}(b) shows that the
MR-curves are close in the shape for all $k_F l$-values, while
the magnitude of the resistivity $\rho$ at low temperature is
varied by more than two orders. This is more clearly seen from
Fig.~\ref{fig2} where
$\Delta\sigma(B)=1/\rho_{xx}(B)-1/\rho_{xx}(0)$ plotted as a
function of reduced magnetic field, $b=B/B_{tr}$. Let us now
analyze the experimental results starting with the case  of high
conductivity.

\subsubsection{High conductivities, $\sigma>20\, G_0$}
\label{ssec:exp1}

\begin{figure}
\includegraphics[width=0.6\linewidth,clip=true]{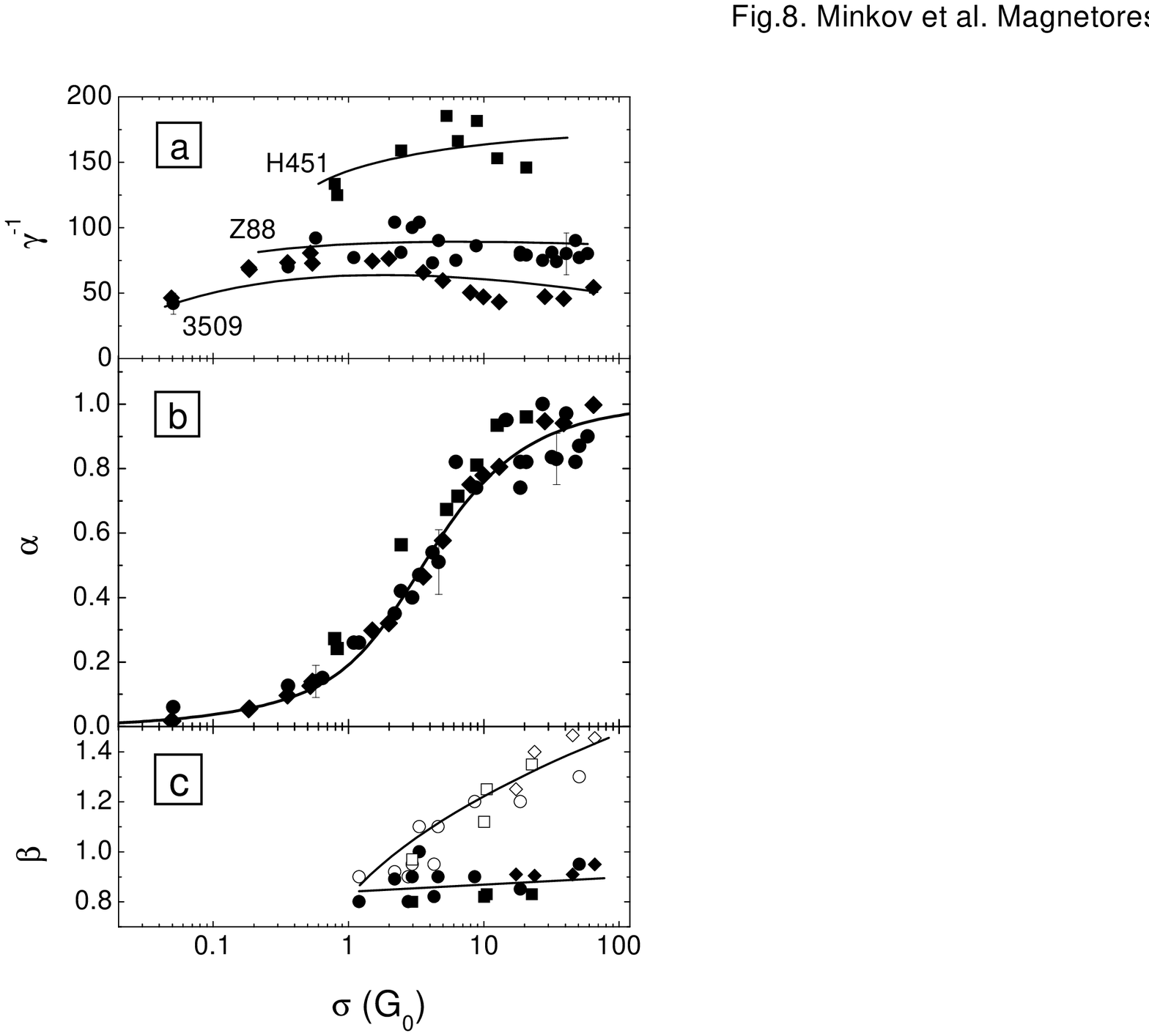}
\caption{The conductivity dependence of the fitting parameters
$\gamma^{-1}$ (a) and prefactor $\alpha$ (b) for structures Z88
(circles), H451 (squares), and 3509 (diamonds), $T=1.5$~K. (c) The
experimental value of the prefactor $\beta$ in the temperature
dependence of the interference quantum correction at $B=0$ (full
symbols) and the slope of the experimental $\sigma$-versus-$\ln\,
T$ dependence (open symbols) as functions of the conductivity at
$T=1.5$~K for structure Z88. Curves in all panels are provided as
a guide for the eye. } \label{fig3}
\end{figure}

In the case of sufficiently high zero-$B$ conductivities,
$\sigma>20\, G_0,$ the value of $k_F l>6$ is large enough and
$\delta\sigma \ll \sigma_0$ in our temperature range. Therefore
the use of WLMC-expression (\ref{eq20}) is really warranted. For
structure Z88, the results of the fit over the magnetic field
range from 0 to 0.25 $B_{tr}$ with $\alpha$ and
$\gamma=\tau/\tau_\varphi $ as fitting parameters are presented in
Fig.~\ref{fig2} by dashed lines (the fit over narrower magnetic
field range gives the close values of the fitting parameters to an
accuracy of $15$\%). The corresponding values of $\alpha$ and
$\gamma$ are given in Table~\ref{tab1}. It is evident that
Eq.~(\ref{eq20}) well describes the experimental data. As seen
from Fig.~\ref{fig3}, where the results of such a data treatment
are collected for all the structures, the prefactor $\alpha$ is
close to unity that agrees with the low value of $\gamma<2\times
10^{-2} \ll 1$. Thus we conclude that the fitting procedure gives
the value of $\tau_\varphi$ which can be directly attributed to
the phase relaxation time.

\begin{figure}
\includegraphics[width=0.8\linewidth,clip=true]{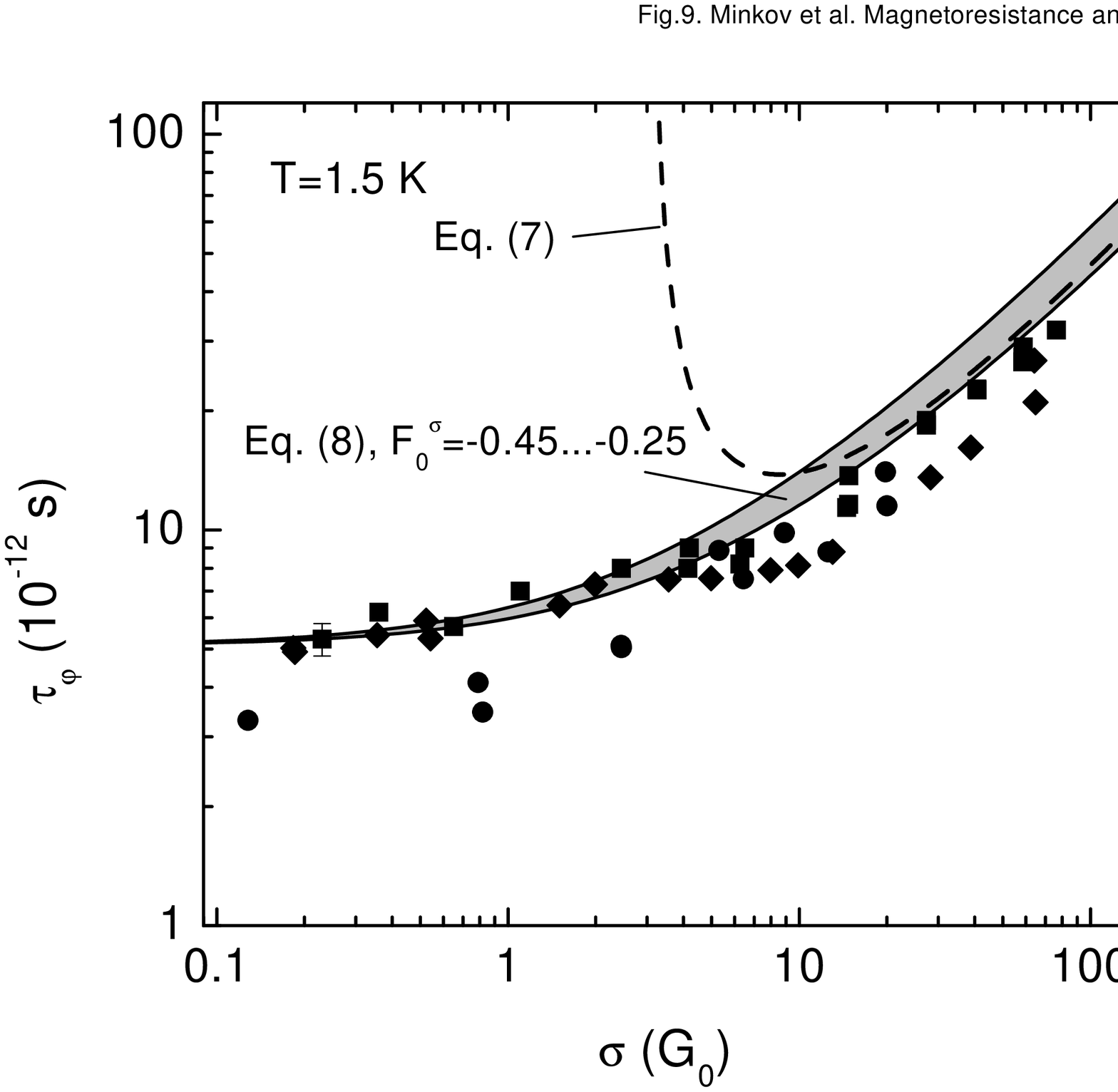}
 \caption
{The conductivity dependence of $\tau_\varphi$ for $T=1.5$ K.
Symbols are the experimental results [designations are the same as
in Fig.~\ref{fig3}(a)]. Dashed line is Eq.~(\ref{eq109}). Shadow
strip represents the solutions of Eq.~(\ref{eq13}) found
numerically for different values of $F_0^\sigma$ from the range
$-0.25$ (upper bounding line) to $-0.45$ (lower one). }
\label{fig4}
\end{figure}

Let us compare the extracted values of the dephasing time with the
theory of the dephasing outlined in Section~\ref{ssec:1}. The
experimental dependences of $\tau_\varphi(\sigma)$ are presented
in Fig.~\ref{fig4}. In the same figure we show the solution of
Eq.~(\ref{eq13}) with $F_0^\sigma$ from the range $-0.45...-0.25$
that corresponds to $K_{ee}=0...0.55$ obtained for the structures
presented here in Ref.~\onlinecite{ourKee}. As seen the
experimental data are in satisfactory agreement with the theory of
Ref.~\onlinecite{Aleiner379}. At first glance, it seems that we
are able to determine the value of $F_0^\sigma$ from
experimentally obtained values of the phase-breaking time.
However, our analysis shows that the total uncertainty in
determination of the phase relaxation time due to the neglect of
the magnetic field dependence of the dephasing rate and due to the
influence of ballistic effects~\cite{our1}  can be estimated as
$20-30$\% that obviously does not allow us to determine
$F_0^\sigma$ by this way reliably. Thus, we assess the dephasing
rate obtained experimentally for high conductivity as agreeing
with the theoretical prediction.

\begin{table}[tbp]
\caption{The parameters for the structures Z88
($T$-dependent quantities are given at T=1.5 K). \label{tab1}}
\begin{ruledtabular}
\begin{tabular}{ccccccc}
$\sigma(G_0)$ & $k_Fl$&$B_{tr} ({\rm Tesla})$ &
$\alpha$ & $\tau_\varphi$
($10^{-12}$~s)
& $\gamma^{-1}$\ & $b_T$\\
 \colrule
50.9           &17.9 &0.029  &0.9   &23.3 &69     & 0.073  \\
18.7           &7.7  &0.12   &0.79   &13.6  & 73   & 0.037  \\
4.66           &2.9  &0.64  &0.53   &9.1   &96    & 0.019 \\
2.2            &2.2   &1.06 &0.35   &7.8  &104    & 0.015 \\
0.575          &1.6   &1.64 &0.14   &5.9   &92     & 0.013 \\

\end{tabular}
\end{ruledtabular}
\end{table}

\subsubsection{Intermediate and low conductivities,
$\sigma<20\,G_0$} \label{ssec:exp2}

Although the WLMC-expression Eq.~(\ref{eq20}) describes the
experimental results rather well (see Fig.~\ref{fig2}), the
correctness of the standard fitting procedure is questionable at
these conductivity values. This is because the prefactor $\alpha$
reveals significant decreasing at $\sigma \lesssim 10\, G_0$ [see
Fig.~\ref{fig3}(b)], implying that the second fitting parameter
$\gamma$ can in principle lose the meaning of the ratio of $\tau$
to $\tau_\varphi$. Therefore it is necessary either to understand
the reasons of such a decrease or to use another theoretical
model. In what follows we will try to employ the results of
Section~\ref{sec:3} to describe the experimental data. We will
show that the decrease of $\alpha$ can be understood within the
framework of the weak localization theory extended to include the
corrections of the second order in $1/g$. This means that the
value of $\gamma_{\rm fit}$ extracted experimentally can be
considered as the true value of $\tau/\tau_\varphi$ down to
$\sigma\simeq 3 G_0$. Remarkably, it turns out that  even in the
case of low zero-$B$ conductivities $0.1\,G_0<\sigma(B=0)<3\,G_0,$
Eq.~(\ref{eq20}) describes the magnetoconductance shape perfectly
(see Fig.~\ref{fig2}). Moreover, surprisingly, this procedure
gives the values of the parameter $\tau_\varphi$, which are close
to that found from Eq.~(\ref{eq13}) down to $0.1\, G_0$ [see
Fig.~\ref{fig4}].

\subsection{Comparison with the self-consistent theory of the MC}

\begin{figure} \includegraphics[width=0.8\linewidth,clip=true]{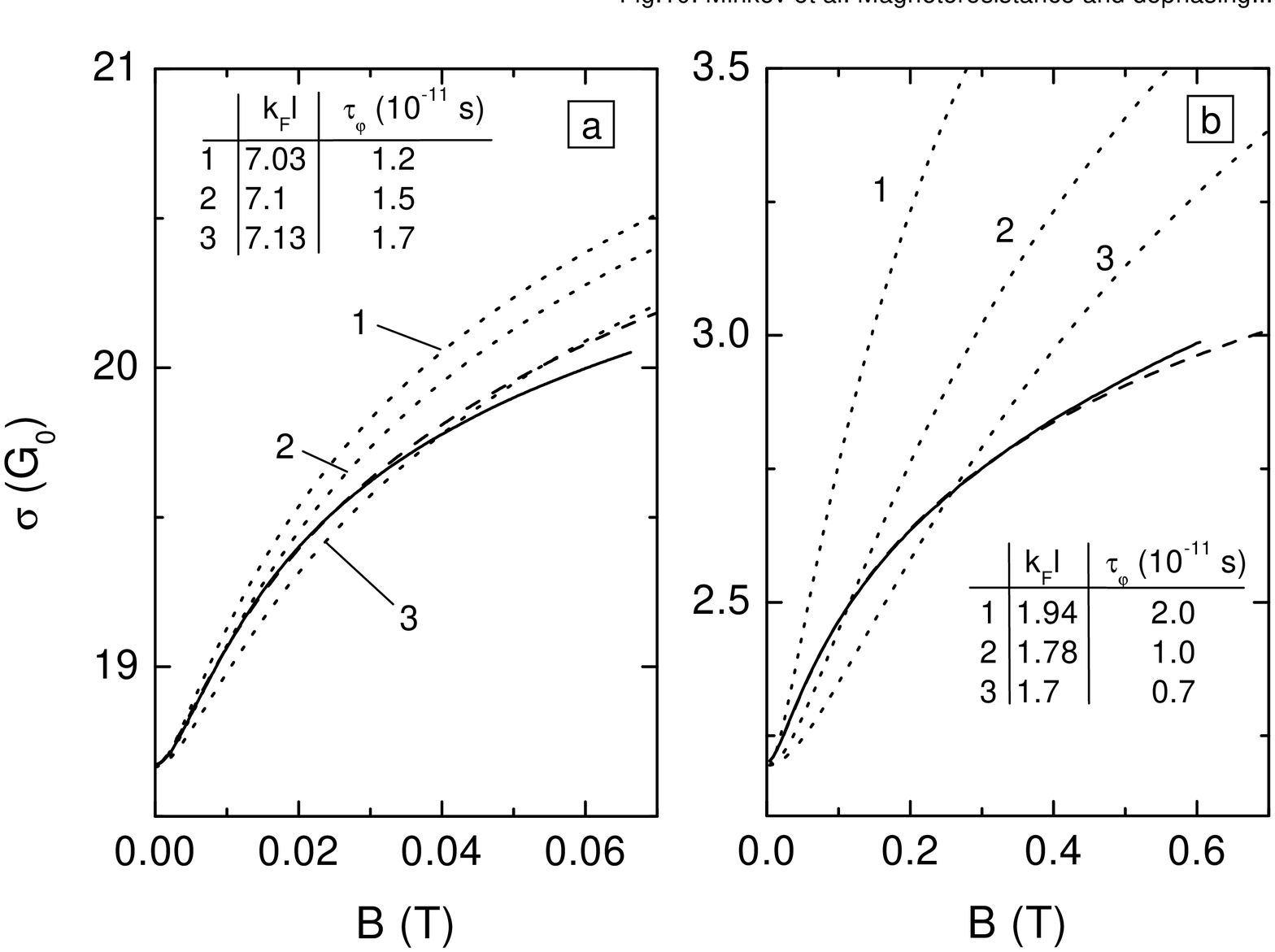}
\caption{The conductivity as a function of magnetic field for
structure Z88 for $\sigma(B=0)=18.7\,G_0$ (a) and $2.2\,G_0$ (b),
$T=1.5$~K. Solid lines are the experimental results. Dotted lines
are solutions of Eq.~(\ref{eq21}) with parameters shown in the
tables, dashed lines are the best fit by Eq.~(\ref{eq20}) with
parameters: $\alpha=0.8$, $\tau_\varphi=1.36\times 10^{-11}$~s,
$B_{tr}=0.12$~T, $k_F l=7.7$ (a) and $\alpha=0.35$,
$\tau_\varphi=0.76\times 10^{-11}$~s, $B_{tr}=0.95$~T, $k_F l=2.2$
(b). } \label{fig5}
\end{figure}

Before applying the approach developed in theoretical part of our
paper to the experimental data, let us use the  self-consistent
Kleinert-Bryksin theory of the Anderson localization in a magnetic
field. According to Ref.~\onlinecite{Bryksin} $\sigma(B)$ is the
solution of the following self-consistent equation

\begin{equation}
  \frac{\sigma}{G_0}=\frac{\sigma_0}{G_0}-\psi\left(\frac{1}{2}+
\frac{l_B^2}{4l^2}+\frac{l_B^2}{4{\cal D} \tau_\varphi}\right)
-\psi\left(\frac{1}{2}+\frac{l_B^2}{4{\cal D} \tau_\varphi}\right),
\label{eq21}
\end{equation}
where ${\cal D}$ itself depends on $\sigma$ as ${\cal
D}=\sigma/(e^2 2\nu)$. The experimental $\sigma$-versus-$B$
dependences together with the solution of Eq.~(\ref{eq21}) for two
values of the conductivity are shown in Fig.~\ref{fig5}. It is
evident that even for the relatively high conductivity
$\sigma(B=0)=18.7\,G_0$ Eq.~(\ref{eq21}) describes the experiment
noticeably worse than Eq.~(\ref{eq20}). In the case of low
$\sigma$ the theory by Kleinert and Bryksin \cite{Bryksin} gives
fully incorrect behavior of $\sigma(B)$. One can try to improve
the expression Eq.~(\ref{eq21}) treating self-consistently not
only the diffusion constant but also the other $\sigma$-dependent
quantities in Eq.~(\ref{eq21}), e.g. using the self-consistent
equation (\ref{eq11}) for $\tau_\varphi$. However, the numerical
calculation shows that this modification of Eq.~(\ref{eq21}) does
not change the results significantly.

\section{Analysis of the experimental results: WL beyond one loop}
\label{sec:exp1}

\subsection{Prefactor in the WLMC-formula}
\label{ssec:pref}

Let us recall now the possible reasons which can in principle lead
to decrease of the prefactor $\alpha$. They were considered in
Section~\ref{ssec:2}. Below we will discuss some of them which
could be relevant in our situation in more detail.

First of all, the decrease of $\alpha$ with decreasing $\sigma$
cannot obviously result from the violation of the diffusion
regime, even for not very high $\sigma,$ because the ratio of
$\tau_\varphi$ and $\tau$ is almost independent of the
conductivity and remains high enough, as illustrated by
Fig.~\ref{fig3}~(a). Also, the range of magnetic fields, where the
fitting of the MC-curves was performed, $b\lesssim 0.25$, does not
include the ballistic range of fields, $B\gtrsim B_{tr}.$

Second, the decrease in the prefactor can in principle result from
the contribution of the e-e interaction in the Cooper channel. It
is apparent that treating the constant $\lambda_0$ in
Eq.~(\ref{eqCEE}) as a fitting parameter, as is usually done, we
are able to describe formally the experimental data by the sum of
Eq.~(\ref{eqEECor}) and Eq.~(\ref{eq20}) with $\alpha=1$. What is
the result? For example, processing the experimental
$\Delta\sigma$-versus-$(B)$ curves for actual conductivity range
($\sigma$ at $T=1.5$~K is about $5\, G_0$) we have obtained that
the value of seed constant $\lambda_0$ changes from
$\lambda_0\simeq 4$ at $T=3$~K to $\lambda_0\simeq -0.62$ at
$T=0.46$~K. Thus, the interaction constant changes with
temperature not only the value but, moreover, the sign. It is
clear that this result is meaningless. If we fix the interaction
constant, say, at the value $\lambda_0=-0.62$ corresponding to the
best fit for $T=0.46$~K, we obtain drastic positive
magnetoresistance for $T=3$~K instead of negative one observed
experimentally. Therefore, already the formal fitting procedure
demonstrates that the electron-electron interaction in the Cooper
channel is not responsible for the decrease of the prefactor with
conductivity decrease under our experimental conditions.

\begin{figure} \includegraphics[width=0.7\linewidth,clip=true]{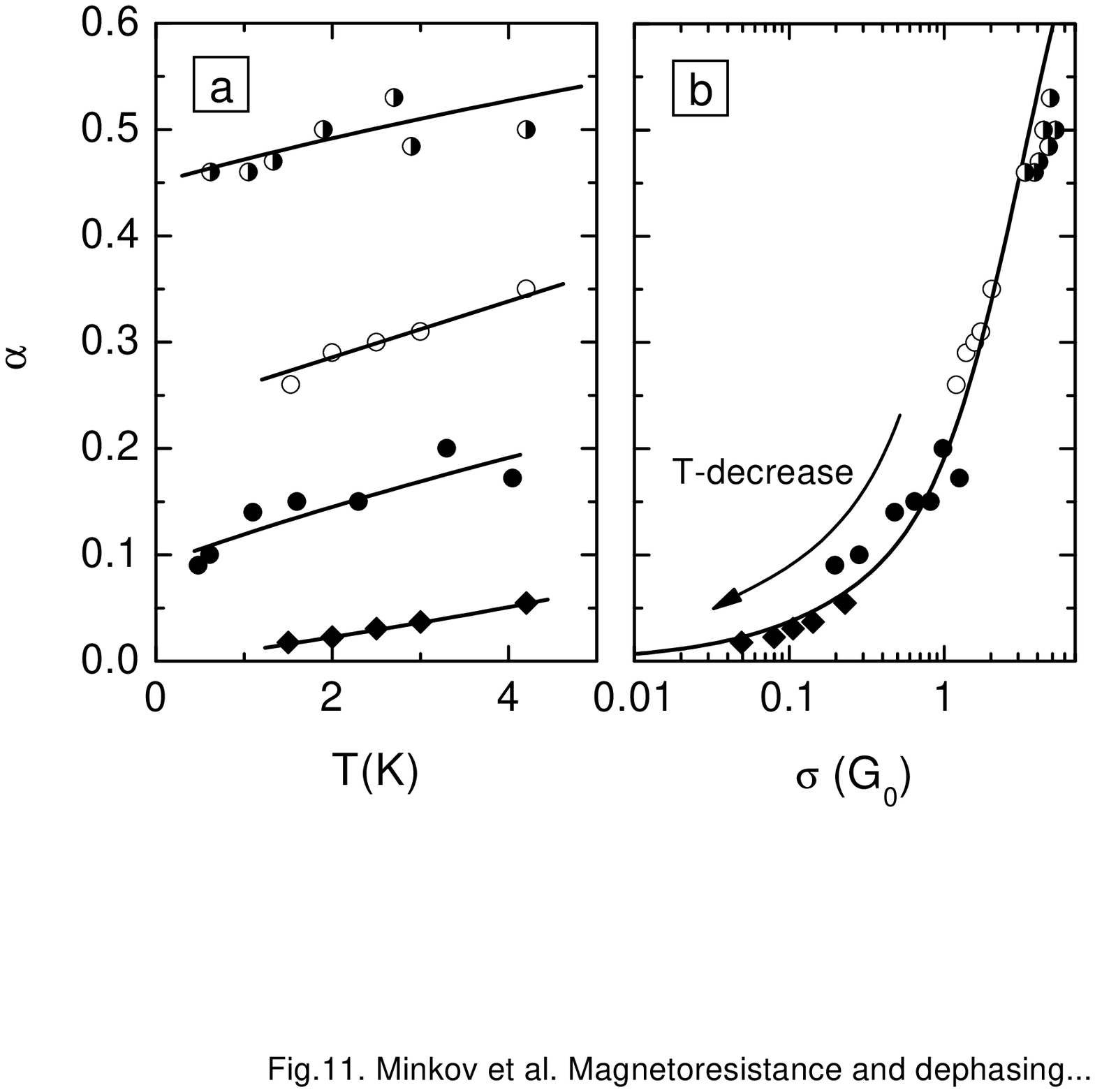}
\caption{The prefactor $\alpha$ plotted against the temperature
(a) and conductivity (b) when the latter changes with temperature.
Circles correspond to structures Z88 when the value of $k_F l$,
controlled by the gate voltage, is $1.6$  (solid circles), $2.0$
(open circles), and $2.9$ (half-filled circles); diamonds are data
for structure 3509 with $k_F l\simeq 1.5$. Solid lines are
provided as a guide for the eye, the line in (b) is just the same
as in Fig.~\ref{fig3}~(b). } \label{fig:alpha(T)}
\end{figure}

However, as discussed in Section~\ref{sec:21}, the actual value of
the effective interaction constant $\lambda_c(T)$ yields a small
corrections to the MC (as an example, see Fig.~\ref{fig:cooper}),
which cannot be responsible for a drastic decrease of $\alpha$.
This is mainly because the ratio $E_F/T\sim 10^2-10^3$ is very
large in structures investigated. Therefore the relevance of the
interaction corrections in the Cooper channel can be ruled out in
the present experiment. Also, we see from Fig.~\ref{fig:alpha(T)}
that temperature dependence of the prefactor is determined by the
$T$-dependence of the conductivity rather than by the
$T$-dependence of the effective interaction in the Cooper channel,
$\lambda_c(T)$,

Let us finally apply the approach described in
Section~\ref{sec:3}. Recall that the lowering of the conductivity
(i) should not change the $T$-dependence of the interference
correction in zero magnetic field [see Eq.~(\ref{eqBeta})] and
(ii) should not influence the shape of the magnetic field
dependence of $\sigma$ leading only to lowering of the prefactor
in dependence $\Delta\sigma(B)$ as given by Eq.~(\ref{alpha2}).

\begin{figure}
\includegraphics[width=0.8\linewidth,clip=true]{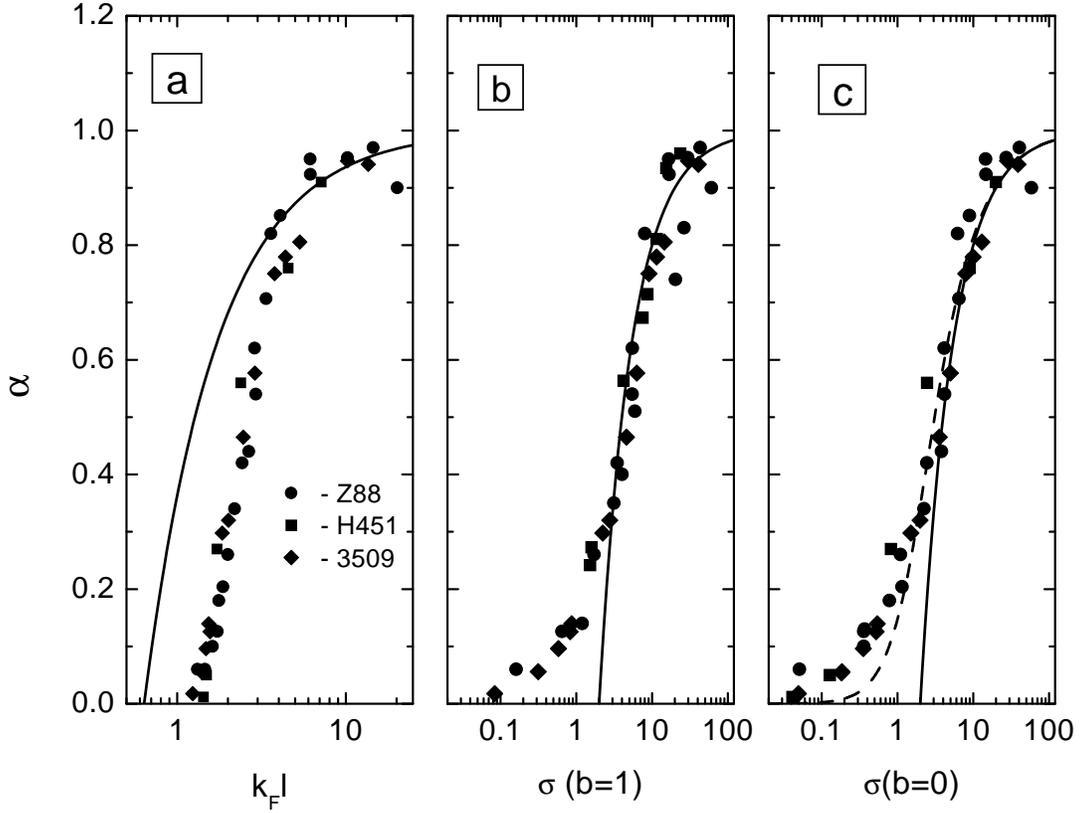}
\caption{The value of prefactor $\alpha$ obtained for $T=1.5$~K as
a function of $k_Fl$ (a), $\sigma(b=1)$ (b), and $\sigma(b=0)$
(c). Solid lines are Eq.~(\ref{alpha2}), dashed line is
Eq.~(\ref{alpha-empirical}) } \label{fig5ab}
\end{figure}

The second point is in a full agreement with our experimental
results. The fitting of the MC was carried out in magnetic fields
up to $b\gg b_T$ for all the curves, and therefore the prefactor
in Eq.~(\ref{eq20}) is determined by the range $T\ll \Omega_B,$
where it is given by Eq.~(\ref{alpha2}). As seen from
Fig.~\ref{fig2}, Eq.~(\ref{eq20}) describes the data perfectly.
The conductivity dependence of the fitting parameter $\alpha$ can
be well described by Eq.~(\ref{alpha2}), as Fig.~\ref{fig5ab}
shows. We see that the second-order perturbative correction to the
prefactor, arising in Eq.~(\ref{eq23}), describes the reduction of
$\alpha$ down to $k_Fl\sim 5$ corresponding to $\alpha \sim 0.8$
[Fig.~\ref{fig5ab}~(a)]. Moreover, as discussed in
Section~\ref{ssec:31} and in Appendix~\ref{ssec:32}, a better
result can be achieved at lower conductivity if one replaces
$\sigma_0$ by $\sigma(b\gtrsim 1)$ obtained for the unitary
ensemble. This is illustrated by Fig.~\ref{fig5ab}(b) in which an
excellent agreement is evident down to $\sigma(b=1)\simeq (2-3)\,
G_0$ corresponding to $\alpha \sim 0.2-0.3$. From a practical
point of view, it is more convenient to use the zero-$B$ value of
the conductivity in Eq.~(\ref{alpha2}). We see from
Fig.~\ref{fig5ab}(c) that this also nicely describes the reduction
of the prefactor $\alpha$, down to slightly higher values of
$\alpha\sim 0.4-0.5$.

An important feature of the prefactor $\alpha$ is that it depends
on the temperature mostly via the $T$-dependence of the
conductivity, as follows from Fig.~\ref{fig:alpha(T)}. Indeed, the
values of the prefactor $\alpha(T)$ obtained for different
temperatures perfectly lie on the same
$\alpha$-versus-$\sigma(b=0)$ curve. We thus see that
Eq.~(\ref{alpha2}) proves to be rather universal. Both the
temperature and the disorder strength affect the value of $\alpha$
only through their influence on the conductivity, so that the
experimental points for different samples, densities, and
temperatures are described by a single $\alpha$-versus-$\sigma$
curve in a broad range of conductivity.

It is tempting to interpret the above universality as an
experimental confirmation of the scaling of the MC with the
magnetic field. Then the conductivity dependence of the prefactor
$\alpha$ might be interpreted as the experimentally determined
$\beta$-function governing the renormalization of the MC. Although
in Appendix~\ref{ssec:32} we have shown that there is no such
scaling in the whole conductivity range (since it is violated in
the third-loop order), an empirical formula resembling those used
for the interpolation of the scaling $\beta$-function between the
WL and SL regimes (see, e.g. Ref.~\onlinecite{AMP99})
\begin{equation}
\alpha\simeq {\sigma \over 2}
\ln\left({1+2/\sigma \over \sqrt{1+4/\sigma^2}}\right)
\label{alpha-empirical}
\end{equation}
appears to describe the prefactor of the MC down to
$\sigma(b=0)\simeq (1-2)\, G_0.$ This can be seen in
Fig.~\ref{fig5ab}(c), where  Eq.~(\ref{alpha-empirical}) is
presented by a dashed curve.

Another prediction of Section~\ref{sec:3} is  that the temperature
dependence of $\sigma$ at $B=0$ which includes both the weak
localization and the electron-electron interaction correction for
the intermediate conductances has to be the same as for the case
$\sigma\gg G_0$,
\begin{equation}
\frac{\sigma(T)}{G_0}= \frac{\sigma_0}{G_0}-
\beta\ln\left(\frac{\tau_\varphi(T)}{\tau}\right) + K_{ee}\ln
\left(\frac{k_BT\tau}{\hbar}\right),
 \label{eq24}
\end{equation}
with $\beta=1$ (if one neglects the corrections in the Cooper
channel). Our measurements show that in the heterostructures
investigated, the temperature dependence of $\sigma$ is actually
logarithmic within the temperature range from $0.45$~K to $4.2$~K
while the value of $\sigma$ remains higher than $(1.0-1.5)\,G_0$,
corresponding to $k_F l\gtrsim 2$. The slope of the
$\sigma$-versus-$\ln{T}$ dependence as a function of $\sigma$ at
$T=1.5$~K is shown in Fig.~\ref{fig3}(c) by open symbols. In order
to obtain the experimental value of the prefactor $\beta$ we have
subtracted from these data the values of $K_{ee}$ which have been
obtained just for the same samples in Ref.~\onlinecite{ourKee}.
The final results are shown in Fig.~\ref{fig3}(c) by solid
symbols. Comparing figures \ref{fig3}(b) and \ref{fig3}(c) one can
see that the prefactor $\alpha$ in MR noticeably deviates down
from unity at $\sigma\simeq (7-8)\, G_0$, whereas the prefactor
$\beta$ in the temperature dependence of $\sigma$ at $B=0$ remains
close to unity down to $\sigma\simeq 1\, G_0$ (deviations from
unity can be attributed to the contribution of the interaction in
the Cooper channel). At lower $\sigma$ it is meaningless to
determine $\beta$, because the temperature dependence of $\sigma$
no longer obeys the logarithmic law.

\begin{figure}
\includegraphics[width=0.6\linewidth,clip=true]{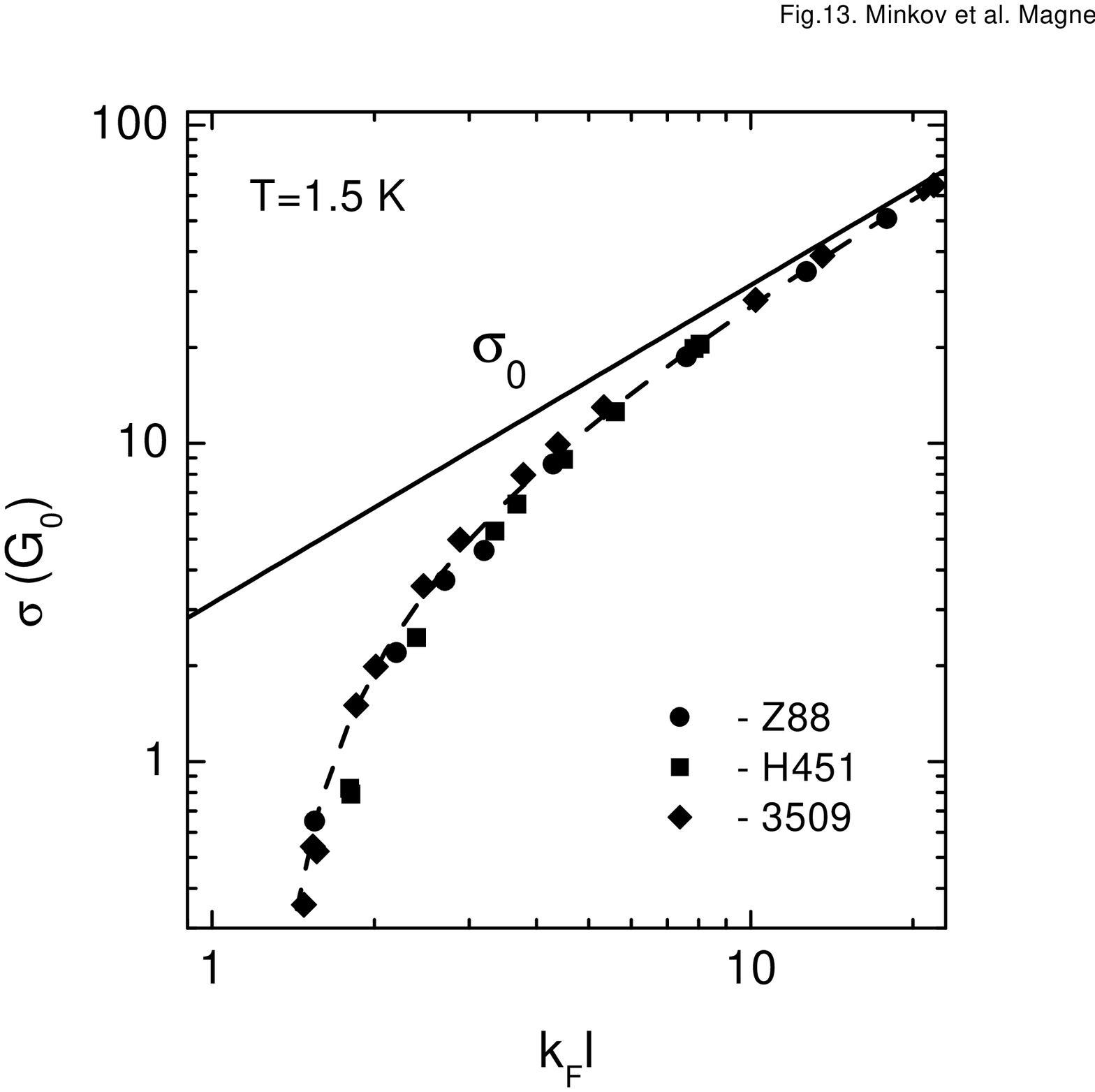}
\caption{The conductivity at T=1.5~K as a function of $k_Fl$. The
value of $k_Fl$ was obtained as described in
Appendix~\ref{sec:kfl}. Dashed line is provided as a guide for the
eye, solid line is $\sigma_0=\pi k_F l G_0$} \label{fig6}
\end{figure}

Figure \ref{fig6}, in which the $\sigma$-versus-$k_F l$ dependence is
plotted, illustrates how strongly the quantum corrections can
suppress the classical conductivity at low temperature. As seen
the value of $\sigma$ is very close to the Drude conductivity at
high $k_F l$ values and significantly less than that at low $k_F
l$. For instance, the ratio $\sigma/\sigma_0$ for $T=1.5$~K is
approximately equal to $0.4$ when $k_F l\simeq 2.5$, so that the
interference and interaction corrections to the conductivity (from
which the first one is the main \cite{WLtoSL}) strongly suppress
the classical conductivity at low temperatures when the parameter
$k_F l$ is small enough.

We arrive at the conclusion that using Eq.~(\ref{eq20}) we obtain
reliably the value of the phase relaxation time with decreasing
the conductivity down to the value of about $3\, G_0$. As seen
from Fig.~\ref{fig4} the values of $\tau_\varphi$ found in this
way demonstrate a good agreement with the dephasing
theory.\cite{Aleiner379} Taking into account terms of the second
order in $1/g$ in the WL theory allows us to understand
quantitatively the magnetic field and temperature dependences of
the conductivity for two-dimensional structures with different
nominal disorder down to the value of the zero-$B$ conductivity
about $\sim e^2/h$. The maximal value of the weak localization
correction reaches $80-90$\% of the Drude conductivity at lowest
temperature, $T=0.45$~K. For the structures investigated this
corresponds to the value of the parameter $k_Fl$ close to two.

\subsection{Temperature dependence of the dephasing rate}
\label{ssec:Tdep}

\begin{figure}
\includegraphics[width=0.9\linewidth,clip=true]{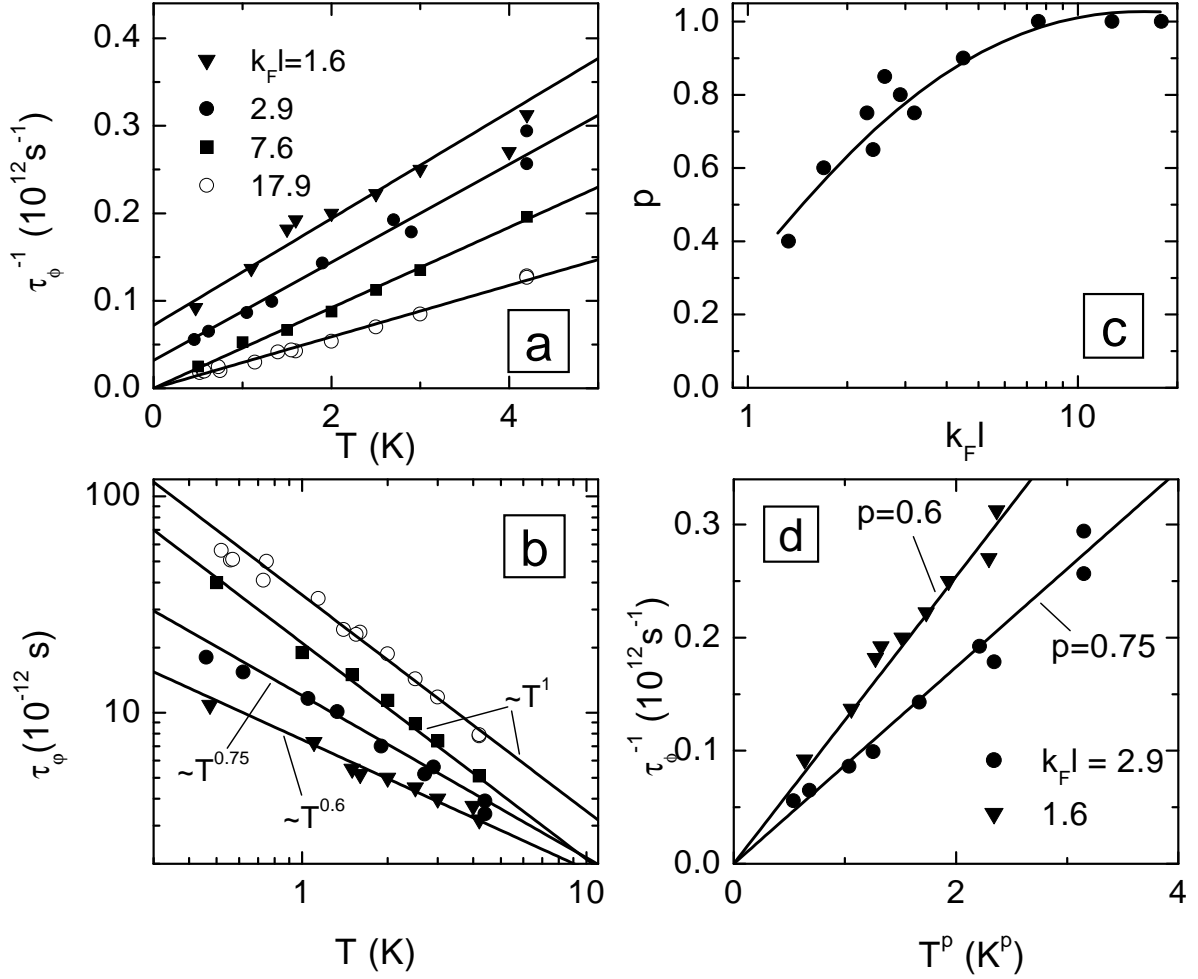}
\caption{(a) The temperature dependence of $\tau_\varphi^{-1}$ for
different values of $k_F l$, structure Z88. (b) The same
experimental data as in (a), presented  in double logarithmic
scale. This plot illustrates that the $T$-dependence of
$\tau_\varphi$ found from the fitting procedure can be
satisfactorily described by the power law. (c) The value of
exponent $p$ obtained from (b) as a function of $k_Fl$ (symbols).
Solid line is a guide for the eye. (d) The values
$\tau_\varphi^{-1}$ plotted against $T^{p}$. Solid lines show
extrapolation to $T=0$. } \label{fig7}
\end{figure}

In this subsection we consider the temperature dependence of
$\tau_\varphi$ extracted from the fitting of the MC by the
WLMC-expression Eq.~(\ref{eq20}). In accordance with the theory
[see equations Eq.~(\ref{eq11}) and Eq.~(\ref{eq13})] we plot the
experimental values of $\tau_\varphi^{-1}$ as a function of $T$ in
Fig.~\ref{fig7}~(a). As seen from this figure, the temperature
dependence of $\tau_\varphi$ can be perfectly described by the
linear-in-$T$ function $\tau_\varphi^{-1}=T/T_0$ and, thus,
$\tau_\varphi$ tends to infinity when $T$ goes to zero, when $k_F
l\gtrsim 5$. At lower values of $k_F l,$ however, a linear
extrapolation of $\tau_\varphi^{-1}$-versus-$T$ dependence gives a
nonzero value of $\tau_\varphi$ at zero temperature. Such a
behavior of $\tau_\varphi$ with temperature, known as phenomenon
of low-temperature saturation of the phase relaxation time, was a
central point of storm discussion in the literature during the
last few years (for the recent review of the problem and for
relevant references see Ref.~\onlinecite{vonDelft}). However, we
demonstrate below that the results presented here have nothing to
do with the saturation of the true dephasing time $\tau_\varphi$
at $T\to 0.$

Let us first follow a standard route and plot our results in
double-logarithmic scale. One can see that the experimental
$T$-dependences of $\tau_\varphi$ (found from
$\tau_\varphi=\tau/\gamma_{\rm fit}$) are well described by the
power law $\tau_\varphi= (T/T_0)^{-p}$ [Fig.~\ref{fig7}~(b)]. The
exponent $p$ is close to unity in wide $k_Fl$-range from $20$ to
$5,$ and slowly decreases when $k_F l$ becomes smaller
[Fig.~\ref{fig7}~(c)]. If we replot the experimental data in the
$\tau_\varphi^{-1}$-vs-$T^{p}$ coordinates, we will see that
$\tau_\varphi^{-1}$ again goes to zero when the temperature tends
to zero [Fig.~\ref{fig7}~(d)]. Thus, the analysis of the
temperature dependence of the fitting parameter $\gamma_{\rm fit}$
shows that the seeming saturation in Fig.~\ref{fig7} can be in
principle explained assuming that the dephasing rate is not
linear-in-$T$ at low enough temperatures and conductances. Indeed,
it looks plausible that the $1/T$-law changes to some
$1/T^{p}$-law with $p<1$ when the parameter $k_F l$ decreases.
Since $\tau_\varphi$ depends on the conductance $\sigma$ which
itself depends on the temperature, the temperature dependence of
$\tau_\varphi$ may be more complicated than simple $T^{-1}$ law.

\begin{figure}
\includegraphics[width=0.6\linewidth,clip=true]{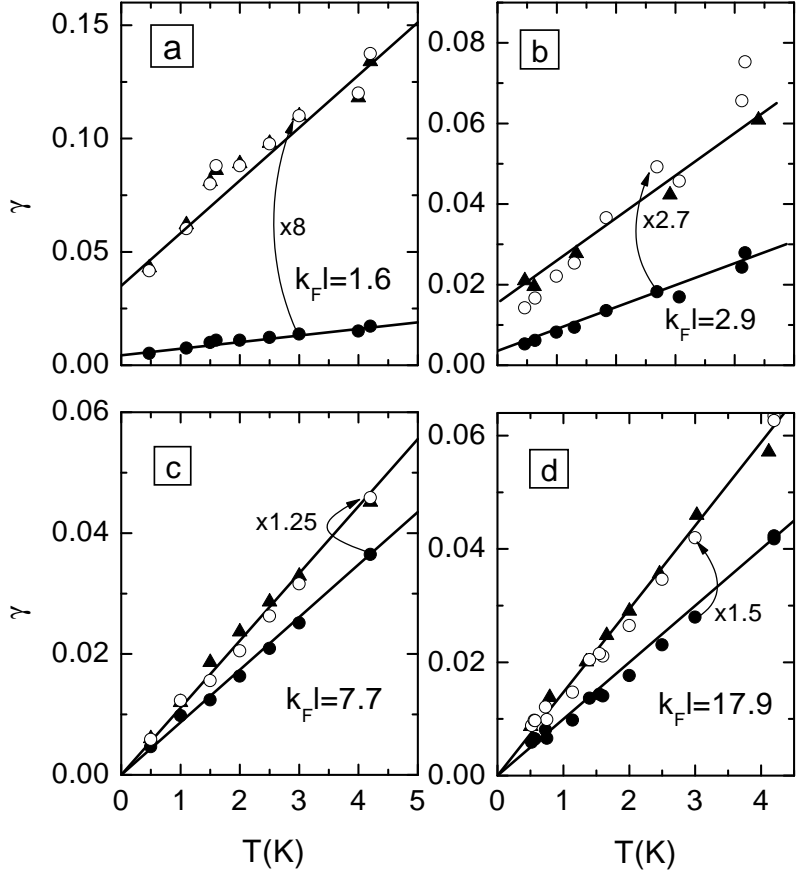}
\caption{ The temperature dependence of the parameter $\gamma$
found from the fit of the magnetoconductivity shape by
Eq.~(\ref{eq20}) (circles), and $\gamma_{\rm fit}$ calculated from
Eq.~(\ref{gamma-sigma}) with ${\cal C}=1/2$ (triangles) with the
use of experimental values $\alpha$, $\sigma(b=0)$ and
$\sigma(b=k_F l/2)$ [as $\sigma(b\gg 1)$] for $k_Fl=1.6$ (a),
$2.9$ (b), $7.7$ (c), and $17.9$ (d), structure Z88. Lines are
provided as a guide for the eye. } \label{fig7aa}
\end{figure}

Already the above consideration shows that our results cannot
serve as the experimental confirmation of the low temperature
saturation of the phase relaxation time. We emphasize, however,
that as discussed above, the experimental value of $\tau_\varphi$
is merely the value of the fitting parameter of magnetoresistance.
It can differ from the true phase relaxation time at $B=0$ when
the conductance in not high, and this is precisely what happens in
our experiment. In particular, this makes it of  a little sense to
analyze the behavior of exponent~$p$. Moreover, as demonstrated in
Section~\ref{ssec:tauphiloc} and in Appendix~\ref{gamma-vs-sigma},
the value of the {\it fitting parameter} $\gamma_{\rm fit}$ {\it
does saturate} when the conductivity at $B=0$ becomes low. This
implies that within the WI regime, the experimentally extracted
value of the dephasing time deviates strongly from the true one.
In fact, the two quantities are close only when
$\sigma(b=0,T)\gtrsim 2 G_0$, otherwise the fitting gives the
information about the localization length instead of the true
dephasing time. In Fig.~\ref{fig7aa} we compare the experimentally
obtained values of $\gamma_{\rm fit}$ with those predicted by
Eq.~(\ref{gamma-sigma}), using the experimental values of the
conductivities and of the prefactor $\alpha$.

Unfortunately, it is not easy to determine experimentally the
strong-$B$ value of the conductivity $\sigma(b\gg 1)$ involved in
Eq.~(\ref{gamma-sigma}), because at strong magnetic fields the
effects that are beyond the WL-theory become important. Therefore
we have chosen to replace $\sigma(b\gg 1)$ by $\sigma(b=k_F l/2)$
to be confident that we are still dealing with the
WL-conductivities. Hence we mistreat partially a $T$-independent
ballistic contribution, so that the factor ${\cal C}$ in
Eq.~(\ref{gamma-sigma}) cannot be determined reliably in this way.
However, the temperature dependence of $\gamma_{\rm fit}$ should
coincide with Eq.~(\ref{gamma-sigma}) up to a numerical factor.
This is clearly seen in Fig.~\ref{fig7aa}. Thus we confirm the
validity of the expression Eq.~(\ref{gamma-sigma}) for the
experimentally extracted value of the dephasing rate. As discussed
in detail in Section~\ref{ssec:tauphiloc}, this expression yields
a rather complicated $T$-dependence of $\gamma_{\rm fit}$ at low
conductances, but this dependence cannot be directly connected
with the $T$-dependence of the true $\tau_\varphi$.

To conclude this subsection, we have shown that the seeming low-$T$ saturation
of the dephasing time at $k_Fl\lesssim 5$ is nothing but an {\it artifact}
of the fitting procedure, which fails to yield the true value of the dephasing
rate at low conductances.
At the same time, the shape of the MC is still perfectly described by the
WLMC-expression at such conductances, but in effect with the localization
length playing a role of the dephasing length.

\subsection{Discussion}
\label{ssec:xi}

\begin{figure}
\includegraphics[width=0.8\linewidth,clip=true]{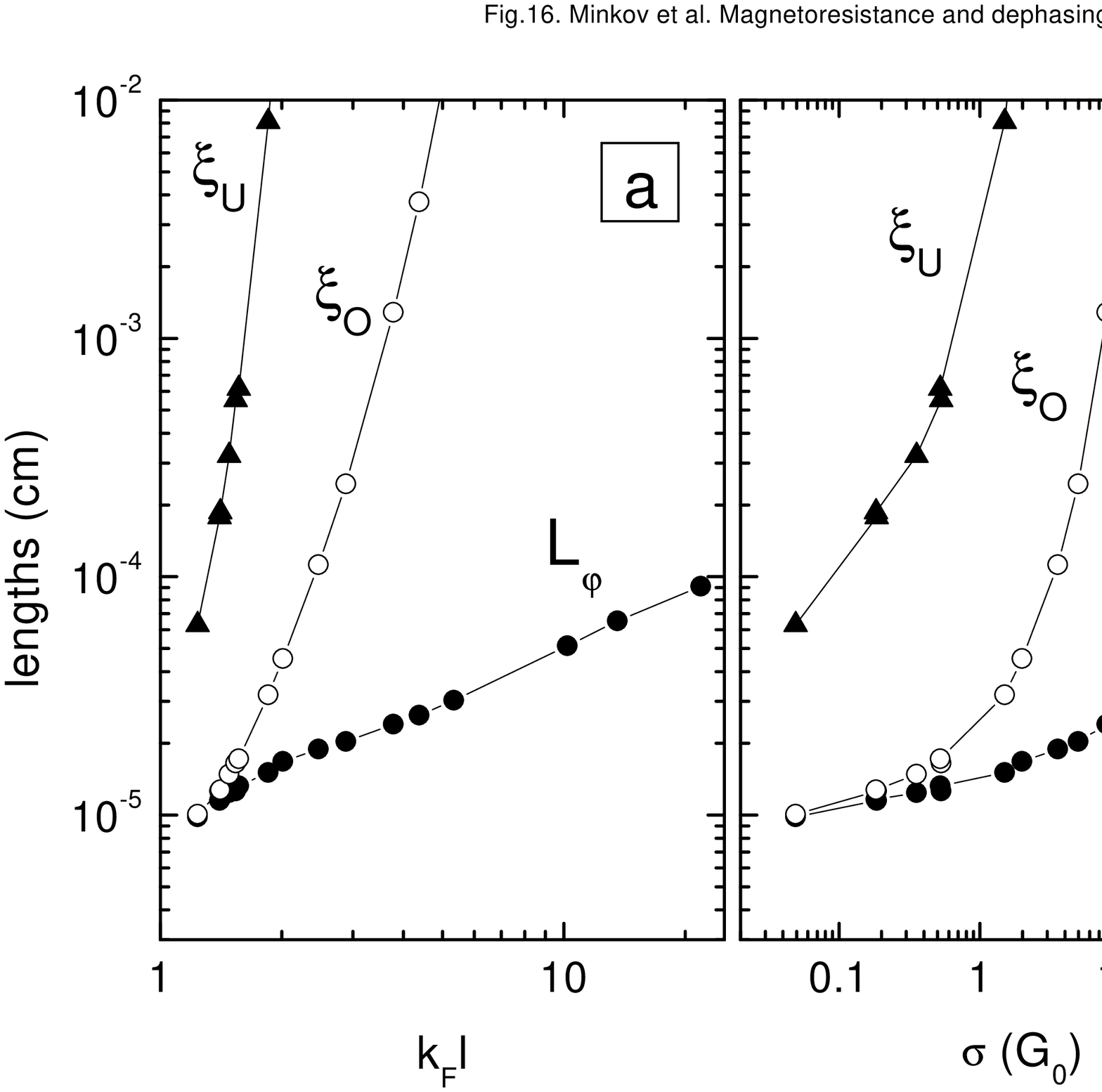}
\caption{The lengths $L_\varphi$, $\xi_O$, and $\xi_U$ as a
function of $k_Fl$ (a) and $\sigma$ for $T=1.5$~K (b). The values
of $\xi_O$ and $\xi_U$ have been calculated from
Eq.~(\ref{eq:XiO}) and Eq.~(\ref{eq:XiU}), respectively, the
length $L_\varphi$ is obtained using the quantity $D$ and
$\tau_\varphi(T=1.5~\text{K})$ obtained experimentally. Structure
Z88.} \label{fig8}
\end{figure}

Now we are in position to understand in what regime the 2D
electron gas considered here is. This is determined by the
characteristic length scales $\xi_O$, $\xi_U$ and $L_\varphi$ as
was considered in the beginning of this paper (see
Section~\ref{sec:int}). As an example, Fig.~\ref{fig8}(a) shows
the relationship between these lengths for structure Z88 as a
function of $k_F l $. The values of $\xi_O$ and $\xi_U$ have been
calculated using Eq.~(\ref{eq:XiO}) and Eq.~(\ref{eq:XiU}),
respectively, whereas the length $L_\varphi=\sqrt{D\tau_\varphi}$
has been found using $D$ and $\tau_\varphi$ obtained
experimentally. It is clearly seen that the value of $L_\varphi$
is always less than both $\xi_O$ and $\xi_U$ when $k_F l\gtrsim2$
and approaches $\xi_O$ with lowering $k_Fl$. This is a direct
manifestation of the fact discussed in
Section~\ref{ssec:tauphiloc}: the experimentally extracted value
of the phase-breaking length saturates with decreasing conductance
at the value determined by the localization length $\xi_O$. This
happens in a narrow range of $1.3<k_Fl< 2$, where the two curves
in Fig.~\ref{fig8} come close to each other. Note, the
low-temperature conductivity in this $k_F l$ range varies
several-fold [see Fig.~\ref{fig8}(b)] reaching values less than
$G_0$. Analogous situation takes place for other heterostructures
investigated.

Thus, we infer that the electron gas in our case is in the WL
regime at $T=1.5$~K for values of $k_Fl$ larger than 2. For lower
values of $k_Fl$ the system is in the WI regime at $T=1.5$~K. We
remind that the magnetoconductivity can still be described by the
WLMC-expression Eq.~(\ref{eq20}) in the WI regime, even though the
conductivity at low temperature becomes less than $G_0$ at low
$k_F l$ values. This holds down to $k_Fl\simeq 1$ which is the
lowest value of $k_Fl$ achieved in the present experiment and thus
addressed in this paper. We therefore conclude that the theory of
the magnetoresistance developed for the SL regime is inapplicable
to our case.

\section{Conclusion}
\label{sec:concl}

We have studied the negative magnetoresistance of a
two-dimensional electron gas in a weak transverse magnetic field
$B$. The analysis has been carried out in a wide range of the
zero-$B$ conductances, including the range of intermediate
conductances (measured in units of $e^2/h$), $g\sim 1$. This range
corresponds to the crossover between the low ($g\ll 1$) and high
($g\gg 1$) conductances. Furthermore, we have considered the
regime of a ``weak insulator'', when the zero-$B$ conductance is
low $g(B=0)<1$ due to the localization at low temperature, whereas
the Drude conductance is high, $g_0\gg 1,$ so that a sufficiently
weak $B$ delocalizes electronic states.

The interpretation of experimental results obtained for
2D electron gas in GaAs/In$_x$Ga$_{1-x}$As/GaAs single quantum
well structures has been based on the theory taking into account terms of
high orders in $1/g$. We have shown that the standard
weak localization theory is adequate for $\sigma\gtrsim 20 G_0$.
Calculating the corrections of the next order in $1/g$ to the MR,
stemming from the interference contribution and from the
mutual effect of WL and Coulomb interaction,
we have expanded the range of the quantitative agreement
between the theory and experiment down to significantly lower
conductances $g\sim 1$.

We have demonstrated that at intermediate conductances the
negative MR is described by the standard WLMC-expression (\ref{eq20}),
with a prefactor $\alpha$ which decreases with
decreasing conductance. We have shown that at not very high $g$
the second-loop corrections dominate over the contribution
of the interaction in the Cooper channel (the Maki-Thompson and
DoS corrections). Thus the second-loop corrections
appears to be the main source of the lowering of
the prefactor, $\alpha=1-2G_0/\sigma$. This formula
describes the experimentally obtained conductivity dependence of
$\alpha$, provided that the fitting is performed in a broad
range of magnetic fields, including those where $\Omega_B\gg T.$
The fitting of the MR allows us to measure
the true value of the phase breaking time within a wide
conductivity range, $\sigma=(3-60)G_0$.
We have shown that the solution
of the equation for $\tau_\varphi$ rather than its first iteration,
describes well the $\tau_\varphi$-versus-$\sigma$ experimental
dependence.

The quantitative agreement between the properly modified WL theory
and experimental results obtained for intermediate conductances
attests that the magnetoconductivity mechanism is unambiguously
diffusive down to $\sigma \simeq 3\, G_0\simeq e^2/h$. Moreover,
an agreement between the extended WL theory and experimental data
persists down to significantly smaller zero-$B$ conductivity
(WI-regime), provided that $k_Fl>1$. In the WI regime, the MR can
be still fitted by the WLMC-formula with a reduced prefactor, but
the experimentally obtained value of the dephasing rate has
nothing to do with the true one.  The corresponding fitting
parameter $\gamma_{\rm fit}$ is determined in the low-$T$ limit by
the localization length and may therefore saturate with at $T\to
0$.

Finally, we have not investigated in detail the magnetoconductivity
in the truly localized (at low-$T$) regime, when both the magnetic length
and the phase-breaking length are greater than
the localization length. The mechanism of a finite-$T$ conductivity in this
situation is not completely clear, when the disorder is weak, $k_Fl\gg 1$.
The experiments on such insulators
reveal features not captured by the conventional ``textbook'' hopping picture.
A thorough study of the magnetoresistance in this regime is therefore of
a great importance for understanding of the low-$T$ transport mechanism in
such ``weakly disordered'' insulators.

\subsection*{Acknowledgment}
We are grateful to I.~L. Aleiner, A.~D. Mirlin, D.~G. Polyakov, P.
W{\"o}lfle, and A.~G. Yashenkin for interesting discussions and
valuable comments. We thank O.~I.~Khrykin, V.~I.~Shashkin, and
B.~N.~Zvonkov  for growing the samples. This work
was supported by the RFBR through Grants No.~02-02-17688,
No.~03-02-16150, and No.~04-02-16626, the Program Russian Science
School 2192.2003.2, the INTAS through Grant No.~1B290,  the CRDF
through Grants No. EK-005-X1 and No. Y1-P-05-11, the Russian
Program {\it Physics of Solid State Nanostructures}, the Program
of Russian Academy of Science, the
Schwerpunktprogramm ``Quanten-Hall-Systeme'', and the SFB195 der
Deutschen Forschungsgemeinschaft.

\appendix

\section{Maki-Thompson correction to the magnetoconductivity}
\label{sec:MT}

In this Appendix we consider in detail the calculation of the
Maki-Thompson correction. To analyze the MC arising due to the
Maki-Thompson correction it is convenient to present this
correction in the concise form,\cite{ayash,ayash1} which is
somewhat different from that used in
Refs.~\onlinecite{Altshuler,AAKL-MIR,Larkin80},
\begin{eqnarray}
\delta\sigma^{\rm MT} &=&
\frac{4 e^2}{\pi \hbar} \int \frac{d^2 {\bf q}}{(2 \pi )^2}
\, \frac{D}{D q^2 + \Omega_B + 1/\tau_\varphi}\nonumber \\
&\times&
\int_{-\infty}^{\infty}\, \frac{d \omega}{2T \sinh^2 (\omega/2T)} \,
\left[ {\rm Im} \Pi_c ({\bf q},\omega,\Omega_B)\right]^2 \,
\left|\Lambda_c({\bf q}, \omega,\Omega_B)\right|^2.
\label{MTayash}
\end{eqnarray}
The functions $\Pi_c$ and $\Lambda_c$ are given by~\cite{Altshuler,Larkin80}
\begin{eqnarray}
\Pi_c ({\bf q}, \omega,\Omega_B) &=&
\ln\left(\frac{2 E_F e^{\bf C}}{\pi T}\right)-\Psi_c({\bf q}, \omega,\Omega_B),
\label{Pic}\\
\Psi_c({\bf q}, \omega,\Omega_B) &=&\psi\left( \frac{1}{2}\left[1+\frac{D q^2 - i \omega +
\Omega_B + \tau_{\varphi}^{-1} }{2\pi T}\right] \right)-\psi\left({1\over 2}\right),
\label{psi} \\
\Lambda_c ({\bf q}, \omega,\Omega_B) &=&  \left[ \frac{1}{\lambda_0}
 + \Pi_c ({\bf q}, \omega,\Omega_B) \right]^{-1}\nonumber \\
 &=&\left[\ln\left(\frac{T_c}{T}\right)-
 \Psi_c ({\bf q}, \omega,\Omega_B)\right]^{-1}
\label{Lambda_c}
\end{eqnarray}
where $\psi$ is digamma function. For simplicity, in
Eq.~(\ref{MTayash}) instead of the summation over quantized
Cooperon momenta we integrate the Cooperon with the mass
$\Omega_B$ over continuous $q$ (which is sufficient for
$\Omega_B\gg \tau_\varphi^{-1}$).

Remarkably, the form Eq.~(\ref{MTayash}) of the Maki-Thompson
correction (in particular the frequency integral appearing there)
is characteristic for the inelastic e-e scattering, see e.g.
Ref.~\onlinecite{GornyiMirlin,ZNA}. The difference is that the
polarization operator $\Pi_c$ and the effective interaction
$\Lambda_c$ are taken in Eq.~(\ref{MTayash}) in the Cooper
channel, instead of the usual particle-hole channel, and the
momentum integral involves a Cooperon. Similarly to conventional
inelastic processes, the imaginary part of $\Pi_c$ comes only from
the ``dynamical'' term $-i\omega$ in Eq.~(\ref{psi}).

At $B=0$, the $q$-integral in Eq.~(\ref{MTayash}) can be split
into two parts, corresponding to small ($Dq^2\ll 2\pi T$) and
large ($Dq^2 \gg 2\pi T$) momenta. In the first contribution, one
can neglect the terms $Dq^2$ and (for $g\gg 1$)
$\tau_\varphi^{-1}$ in the function $\Psi_c$. Then the
$q$-integral yields a logarithmic factor $\ln(T\tau_\varphi)\sim
\ln g$. Furthermore, one can replace $\Lambda_c$ by
$\lambda_c(T)$. Performing then the frequency integral, we arrive
at Eq.~(\ref{MTB=0}), in agreement with
Ref.~\onlinecite{Larkin80}.

When calculating the contribution of large momenta, $Dq^2>2\pi
T$,\ one can use the asymptotics of digamma function at large
argument. Then the $q$-integral, having the structure $\int (dx/x)
(T/x)^2 \ln^{-2}(T_c/x)$ with $x=Dq^2$, is determined by the lower
limit $x\sim 2\pi T$. The result of integration is also
proportional to $\lambda^2_c(T)$, but in contrast to the
contribution of small momenta, this integral does not produce a
logarithmic term $\ln T\tau_\varphi.$ Therefore this contribution
can be neglected at $g\gg 1$;\cite{Larkin80} however, with
decreasing $g$ the two contributions become comparable.

In a finite magnetic field, the structure of
Eqs.~(\ref{MTayash}--\ref{Lambda_c}) suggests that the behavior of
the Maki-Thompson correction depends on the value of the parameter
$\Omega_B/2\pi T$. For $\Omega_B\ll 2\pi T$ the contribution of
small $q$ yields the MR given by Eq.~(\ref{MRMT-weak}). The
contribution of large $q$ to the MC depends only weakly on $B$ for
any $g$, since to the leading order in $\Omega_B/2\pi T$ this
contribution to $\delta\sigma^{\rm MT}$ is $B$-independent.

For $\Omega_B\gg 2\pi T$, one can again use the asymptotics
of digamma function at large argument (now for the arbitrary momenta).
The momentum integral is then determined by $Dq^2\sim \Omega_B,$
yielding
\begin{equation}
\delta\sigma^{\rm MT}\propto
\left(\frac{2\pi T}{\Omega_B}\right)^2\frac{1}{\ln^2(2\pi T_c/\Omega_B)},
\qquad \Omega_B\gg 2\pi T.
\label{MT-strongB}
\end{equation}
Therefore the Maki-Thompson contribution to the MC,
$\Delta\sigma^{\rm MT}(B)=\delta\sigma^{\rm
MT}(B)-\delta\sigma^{\rm MT}(0),$ saturates in the limit of high
$B$ at $\Delta\sigma^{\rm MT}=-\delta\sigma^{\rm MT}(0),$ where
$\delta\sigma^{\rm MT}(0)$ is given by Eq.~(\ref{MTB=0}). This
makes it possible to describe the behavior of the Maki-Thompson
correction to the MC by Eq.~(\ref{eq20a}) in the whole range of
magnetic fields, with $\alpha_{\rm MT}=-\pi^2\lambda^2_c(T)/6$ and
the replacement $1/\tau \to 2\pi T$,
\begin{equation}
\Delta\sigma^{\rm MT}(B)
\simeq -{\pi^2\lambda^2_c(T) \over 6}{\cal H}(b/2\pi T\tau,1/2\pi T\tau_\varphi).
\label{MT-Hik}
\end{equation}
Although at $\Omega_B\gg T$ this expression gives the asymptotics
different from Eq.~(\ref{MT-strongB}), the precise way of the
saturation of the Maki-Thompson contribution to the MC is
irrelevant, since at $\Omega_B\gg T$ the DoS-term dominates
$\Delta\sigma_{ee}^{\rm C}$ there, see Section~\ref{sec:3}.

\section{Scaling of the conductance in the cross-over between
the unitary and orthogonal ensemble}
\label{ssec:32}
\vspace{0.1cm}

One might be tempted to reformulate the results of Section IVA
in terms of the renormalization group (RG) equation for the MC.
That is, one might conjecture the existence of a
single-parameter scaling of the MC with the
magnetic field in the cross-over
between the unitary and orthogonal ensembles.
Indeed, the second-loop perturbative correction to the MC
is logarithmic in $B$ which might correspond
to the scaling of the MC with $l_B$,
governed by the ``cross-over''
$\beta$-function,
\begin{equation}
\beta_{UO}=\beta_O-\beta_U=-{1\over {\mathsf g}}+{1\over 2 {\mathsf g}^2}, \qquad {\mathsf g}\gg 1.
\label{beta-cross}
\end{equation}
In other words, given the value of the conductivity at $B=B_{tr}$
[Eq.~(\ref{dsigma2D})], one could be able to restore the MC for a
fixed $T$ in the wide range of weaker magnetic fields ($l_B\ll
L_\varphi$) including those where $\sigma(B,T)<G_0$, using a
single RG-equation. Starting at $b\simeq 1$ from a high
conductance, ${\tilde {\mathsf g}}_0(L_\varphi)={\mathsf
g}_0-(1/2{\mathsf g}_0)\ln(L_\varphi/l)\gg 1,$ one would obtain
the following expression for the renormalized conductance at
weaker $B$,
\begin{eqnarray}
{\mathsf g}(l_B,L_\varphi)&\simeq &{\tilde {\mathsf g}}_0(L_\varphi)-
\left(1-{1\over 2{\tilde {\mathsf g}}_0(L_\varphi)}\right)\ln\left({l_B\over l}\right)
\label{RG1} \\
&=& {\mathsf g}_0-{1\over 2 {\mathsf g}_0}\ln \left({L_\varphi\over l}\right) -
\left(1-{1 \over 2 {\mathsf g}_0}\right)\ln \left({l_B\over l}\right) + {\cal O}(1/{\mathsf g}_0^3),
\label{RG2}
\end{eqnarray}
which agrees with the result of the perturbation theory.

However, an analysis of higher order corrections with the help of
Eqs.~\ref{betaU} and \ref{betaO} shows that there is no true
scaling of the MC with $l_B.$ This can be seen in the third-loop
order, implying that an ``approximate scaling'' with $l_B$ takes
place only when the conductance is sufficiently high. On the other
hand, the conventional scaling with the system size (in the
present case -- with the phase-breaking length $L_\varphi$) is
applicable to the unitary-orthogonal cross-over.\cite{Lerner-Imry}
To illustrate this in the nontrivial case, when the localization
length becomes $B$-dependent (see Section~\ref{sec:int}), we fix
the magnetic length to be the shortest macroscopic scale. We also
set the value of the phase-relaxation length to lie between the
two localization lengths,
\begin{equation}
l\ll l_B \ll \xi_O \ll L_\varphi \ll \xi_U.
\label{RG3}
\end{equation}
In this situation, electrons are localized at $B=0$, but
the magnetic field is chosen to delocalize the electronic states.

We start the scaling procedure at the microscopic scale, $L=l,$ where
${\mathsf g}\simeq {\mathsf g}_0\gg 1$ and increase $L$ up to $L=l_B$.
The renormalization on such scales ($L<l_B$) is governed
by the orthogonal $\beta$-function (\ref{betaO}), and we get
\begin{equation}
{\mathsf g}(L)={\mathsf g}_0-\ln\left({L\over l}\right)+
{\cal O}\left({\ln(L/l)\over {\mathsf g}_0^3}\right).
\label{RG4}
\end{equation}
At scales larger than $l_B$, the scaling is governed by
the unitary $\beta$-function (\ref{betaU}).
The starting conductance is given by
${\mathsf g}(L=l_B)={\mathsf g}_0-\ln(l_B/l)\gg 1$ and we find (for brevity, we measure the
lengths in units of $l$ below),
\begin{eqnarray}
{\mathsf g}(L)&=&{\mathsf g}(l_B)-{1\over 2 {\mathsf g}(l_B)}\ln\left({L\over l_B}\right)+
{\cal O}(1/{\mathsf g}^3)
\nonumber \\
&\simeq& {\mathsf g}_0-{1\over 2 {\mathsf g}_0}\ln L -
\left(1-{1 \over 2 {\mathsf g}_0}\right)\ln l_B-
{1\over 2 {\mathsf g}_0^2}
\ln l_B \left[\ln L - \ln l_B\right].
\label{RG5}
\end{eqnarray}

The first three terms in this expression coincide with
Eq.~(\ref{RG2}). The term containing the factor $1-1/2{\mathsf
g}_0$ in Eq.~(\ref{RG5}) leads to the decrease of the prefactor
$\alpha$ and has been already discussed above. One can see,
however, that the last term $\sim {\cal O}(1/{\mathsf g}_0^2)$ in
Eq.~(\ref{RG5}) appears to violate the scaling of the MC with
$l_B,$ as seen from the comparison of Eq.~(\ref{RG5}) and
Eq.~\ref{RG2}: this term is absent in Eq.~(\ref{RG2}). Note that
this term cannot be canceled out by higher terms in the expansion
of $\beta$-functions because the $1/{\mathsf g}^3$-terms vanish in
both Eq.~(\ref{betaO}) and Eq.~(\ref{betaU}). Taking a derivative
of ${\mathsf g}(L,l_B)$ with respect to $l_B$, one finds that it
does not depend solely on the value of ${\mathsf g}(L,l_B)$
itself. In particular, the term of the type ${\mathsf g}_0^{-2}\ln
L$ arises, meaning the failure of the conjectured scaling equation
Eq.~(\ref{beta-cross}) in the ${\cal O}(1/{\mathsf g}^2)$ order.

The above standard scaling procedure allows one to estimate the localization
length $\xi_{UO}$ at finite $B$ from the equation
$g(\xi_{UO})\simeq 0$, yielding
\begin{equation}
\xi_{UO}(B)\sim l_B \exp\left([\pi k_F l/2 -\ln(l_B/l)]^2\right),
\label{xiUO}
\end{equation}
in agreement with Ref.~\onlinecite{Lerner-Imry}.

Thus, while the scaling of the magnetoconductance with $B$ does
not exist at low renormalized conductance, one can nevertheless
use the formula Eq.~(\ref{RG2}) with ${\tilde {\mathsf g}}_0$
found at high fields $b\gtrsim 1$ in a rather wide range of
fields. This implies that the effective prefactor $\alpha$ found
from the fitting of the MC by the WL expression can be
approximated by
\begin{equation}
\alpha_{WL}\simeq 1-{G_0\over \sigma(b\simeq 1,T)},
\end{equation}
when the interaction corrections are neglected.

\section{Dephasing rate extracted from WLMC-formula}
\label{gamma-vs-sigma}

In this Appendix we derive a general expression for
the experimentally extracted value of $\gamma_{\rm fit}$.
We start with the reiteration of the scheme used
for the derivation of the WLMC-expression
and its generalization including the second-loop terms.
For simplicity, we will not consider the contribution
of $\delta\sigma_2^{\rm{I}\times {\rm WL}}$ in this Appendix and
concentrate on the interference corrections.
The MC $\Delta\sigma(b)$ is defined as
\begin{equation}
\Delta\sigma(b)=\sigma(b)-\sigma(0).
\label{MC-definition}
\end{equation}
This definition is obviously
more general than $\Delta\sigma(b)=\delta\sigma(b)-\delta\sigma(0),$
frequently used for $g\gg 1,$ i.e. when the conductivity
corrections are small, compared to the Drude conductivity $\sigma_0$.

For $g\gg 1$ it is sufficient to take into account
only the one-loop WL corrections
to $\sigma_0$ (below we measure the conductivities in units of $G_0$),
\begin{eqnarray}
\sigma(b)&=&\sigma_0+\delta\sigma(b)\simeq \sigma_0+
\psi\left({1\over2}+{\gamma \over b}\right)
- \psi\left({1 \over 2}+{1\over b}\right)+\delta\sigma_{\rm ball}, \quad b\ll 1,
\label{sigmab} \\
\sigma(0)&=&\sigma_0+\delta\sigma(0)=
\sigma_0+\ln\gamma+\delta\sigma_{\rm ball}.
\label{sigmab=0}
\end{eqnarray}
Here $\delta\sigma_{\rm ball}$ accounts for the non-logarithmic
contributions of the ballistic interfering paths with lengths
$L\lesssim l$ and the non-backscattering processes.\cite{dmit} In
zero $B$ (and actually for $b\ll 1$) this contribution for the
case of white-noise disorder (short-range impurities)
reads~\cite{dmit}
\begin{equation}
\delta\sigma_{\rm ball}(b=0)=\ln 2.
\label{ballisticpaths}
\end{equation}
Using Eqs.~(\ref{MC-definition}), (\ref{sigmab}), and
(\ref{sigmab=0}), we see that  the terms $\sigma_0$ and
$\delta\sigma_{\rm ball}$ from (\ref{sigmab}) and (\ref{sigmab=0})
cancel out in Eq.~(\ref{MC-definition}), so that the one-loop
magnetoconductivity $\Delta\sigma_1(b)$ is determined solely by
the logarithmic conductivity corrections,
\begin{equation}
\Delta\sigma_{(1)}(b)=\psi\left({1\over2}+{\gamma_b \over b}\right)
- \psi\left({1 \over 2}+{1\over b}\right)-\ln\gamma,\quad b\ll 1
\label{MC-oneloop}
\end{equation}
which is just the standard WLMC-expression with $\alpha=1$.

It is worth emphasizing that the value of $\gamma$ extracted from
the experiment with the use of Eq.~(\ref{MC-oneloop}) is mostly
determined by $\gamma$ from the last term ($\ln\gamma$) in
Eq.~(\ref{MC-oneloop}), i.e. it comes from $\delta\sigma(b=0)$.
This can be seen from the asymptotics Eq.~(\ref{strBWLMR}) of the
MC at strong $B$. In particular, when the MC is logarithmic, the
value of $\gamma_b$ entering into the first digamma function (i.e.
related to the dephasing in a finite $B$, hence the subscript $b$)
appears only in the subleading term ${\cal O}(\gamma/b)$, see
Ref.~\onlinecite{Aleiner379}. This difference is, however,
unimportant for the one-loop correction to the MC, but it becomes
essential with lowering $g$.

Now we rewrite Eq.~(\ref{MC-oneloop}) in a slightly different form
\begin{eqnarray}
\Delta\sigma_{(1)}(b)&=&\delta\sigma(b)-[\sigma(0)-\sigma_0] \nonumber \\
&\simeq& \psi\left({1\over2}+{\gamma_b \over b}\right)
- \psi\left({1 \over 2}+{1\over b}\right)+\delta\sigma_{\rm ball}-
[\sigma(0)-\sigma(b\gg 1)].
\label{rewrittenMC1}
\end{eqnarray}
Here we have used that $\sigma(b\gg 1)\simeq \sigma_0$
within the one-loop approximation, since the ballistic contribution
is also suppressed by a strong magnetic field,
$\delta\sigma_{\rm ball}(b\gg 1)\to 0$.
We thus see that instead of $\ln \gamma$
a structure, expressed in term of conductivities,
$\sigma(0)-\sigma(b\gg 1)-\delta\sigma_{\rm ball},$ appears.

When the second-loop perturbative contribution is included,
we have an analogous expression, see Section~\ref{ssec:31}
\begin{eqnarray}
\Delta\sigma_{(1+2)}(b)&=&\left(1-\frac{1}{\sigma_0}\right)\left\{
\psi\left({1\over2}+{\gamma_b \over b}\right)
- \psi\left({1 \over 2}+{1\over b}\right)\right\}+
\delta\sigma_{\rm ball}\nonumber \\
&-&[\sigma(0)-\sigma(b\gg 1)].
\label{rewrittenMC2}
\end{eqnarray}
The zero-$B$ conductivity, Eq.~(\ref{sigmab=0}), remains
unchanged, while the conductivity at strong $B$ is now
renormalized by the second-loop diffuson correction,
Eq.~(\ref{dsigma2D}), $\sigma(b\gg 1)\simeq
\sigma_0+(1/\sigma_0)\ln\gamma.$ Recalling that $\alpha_{\rm
WL}=1-1/\sigma_0,$ we get
\begin{eqnarray}
\Delta\sigma_{(1+2)}(b)&=&\alpha_{\rm WL}\left\{
\psi\left({1\over2}+{\gamma_b \over b}\right)
- \psi\left({1 \over 2}+{1\over b}\right)\right.\nonumber \\
&+&\left.\frac{1}{\alpha_{\rm WL}}\Big[\sigma(b\gg 1)-
\sigma(0)+\delta\sigma_{\rm ball}\Big]\right\},
\label{rewrittenMC2a}
\end{eqnarray}
implying that the role of $\ln\gamma$ in the generalized
WLMC-expression is played by the combination in the last term in
Eq.~(\ref{rewrittenMC2a}),
\begin{equation}
\ln\gamma_{\rm fit} \to
-\frac{1}{\alpha_{\rm WL}}\Big[\sigma(b\gg 1)-
\sigma(0)+\delta\sigma_{\rm ball}\Big].
\label{newgamma}
\end{equation}
Thus the experimentally extracted value of the parameter $\gamma_{\rm fit}$
is given by
\begin{equation}
\gamma_{\rm fit}={\cal C}_{\rm ball}
\exp\left\{{1\over \alpha_{\rm WL}}\Big[\sigma(0)-\sigma(b\gg 1)\Big]\right\},
\label{gamma-sigmaWL}
\end{equation}
where ${\cal C}_{\rm ball}$ is determined by the ballistic
contribution $\delta\sigma_{\rm ball}$ which depends on the character
of disorder. For point-like impurities,
${\cal C}_{\rm ball}\simeq 1/2$.

It turns out that under the condition $\sigma(b\gg 1)\gtrsim 3 G_0$
(when the second-loop
expression for $\sigma(b)$ is sufficient),
this expression is valid for arbitrary $\sigma(b=0),$
including $\sigma(b=0)\ll G_0$, i.e. in the WI regime.

\section{Experimental determination of the Drude conductivity}
\label{sec:kfl}

In this Appendix we describe how the values of $B_{tr}$ and $k_F l,$
playing a pivotal role in the theory of weak localization, can be obtained
experimentally. Since $B_{tr}$ is expressed as
$B_{tr}=\hbar/(2el^2)$ where $l$ is elastic mean free path, it can
be found from the Drude conductivity $\sigma_0=\pi k_F l$ and
electron concentration $n=k_F^2/(2\pi)$. The electron
concentration can be obtained from the Hall effect. The question
is: how can one obtain $\sigma_0$?

In the case of relatively high
conductivity the experimental low-temperature value of $\sigma$
instead of $\sigma_0$ is often used. Let us estimate an error for
the concrete case of $\sigma_0=20\,G_0$ and
$\gamma=\tau/\tau_\varphi=0.01$. The use of
$\sigma=\sigma_0+G_0\ln{\gamma}\simeq 15.4\,G_0$ (the interaction
correction is neglected) instead of $\sigma_0=20\, G_0$ for
determination of $B_{tr}$ leads to overestimation of $B_{tr}$ by a
factor of about $1.4$. In its turn this results in an
overestimation of $\tau_\varphi$ found experimentally by just the
same factor.
Another possibility is to measure the high-$T$ value of the
conductivity, believing that the quantum corrections are destroyed
by temperature. However, at high temperatures the electron-phonon
scattering comes into play. Also, the ballistic contribution~\cite{ZNA}
of the e-e interaction is $T$-dependent, thus making it difficult
to determine the true value of $\sigma_0$.

In this paper the value of  the Drude conductivity has been
obtained by subtraction of the both interference and interaction
corrections from the experimental value of the conductivity at
$B=0$, using Eq.~(\ref{eq24}). The value of $K_{ee}$ was measured
for the structure presented here in Ref.~\onlinecite{ourKee}. It
has been found that the value of $K_{ee}$ is about $0.3-0.5$ at
$k_Fl\gtrsim 5$ and decreases with $k_F l$-decrease, almost
vanishing at the lowest $k_Fl \simeq 2$. The right-hand side of
Eq.~(\ref{eq24}) itself depends on the value of
$\tau/\tau_\varphi$, which can be found from the negative
magnetoresistance. Therefore, treating the interference induced
negative magnetoresistance we used the successive approximation
method applying both equations Eq.~(\ref{eq20}) and
(Eq.~\ref{eq24}). For the first approximation we set $\sigma_0$
equal to $\sigma$, found $B_{tr}$ and, then, determined $\gamma$
from the fit of magnetoresistance by Eq.~(\ref{eq20}) using it and
$\alpha$ as fitting parameters. After that we substituted $\gamma$
into Eq.~(\ref{eq24}) and found the corrected value of $\sigma_0$
and so on. An output of this procedure is the value of the
prefactor $\alpha$, the ratio $\tau/\tau_\varphi=\gamma$ and the
value of the Drude conductivity $\sigma_0$. It is sufficient to
make from five to eight iterations to achieve an accuracy in the
determination of $\sigma_0$ and $\gamma$ better than 10\%. So
complicated method is not significant for high conductivity
($\sigma \gtrsim 50 G_0$) when the quantum corrections are
relatively small. It should be noted that in the above procedure
we do not take into account the non-backscattering contribution to
the weak localization, which gives an additional temperature
independent positive interference contribution equal to $G_0
\ln\,2$.\cite{dmit} In view of this and all other facts not
pointed out here, we estimate an error in determination of
$\sigma_0$ to be about $\pm 0.5\, G_0$.


\begin{thebibliography}{33}
\bibitem{foot-Lvstau} Strictly speaking, in this regime the phase-breaking
length loses its meaning, since for electrons localized at $B=0$ the
largest spatial scale is given by $\xi_O$. In this sense, it is
meaningful to use the quantity $\tau_\varphi$ instead of $L_\varphi$.
Then one can formally define $L_\varphi$ through $\tau_\varphi$
using the relation valid for delocalized electrons. It is such
phase-breaking length which we use when the SL and WI regimes are
considered.
\bibitem{foot-nonpert} It is worth noting that at
$\sigma\sim (2-3)G_0,$ contributions
$\propto \exp(-\sigma/G_0)$ to the MC, non-perturbative in $1/g$,
are still very small and can be neglected.
We do not consider such terms in this paper.
\bibitem{foot-time} A similar situation occurs for the
interaction correction to the conductivity in the
ballistic regime ($T\tau\agt 1$), see Ref.~\onlinecite{ZNA},
where the correction to the
conductivity is determined by the $T$-dependent renormalization of the impurity
scattering cross-section (and hence of the transport mean free time)
by Friedel oscillations.
\bibitem{foot-AL} Here we neglect the Aslamazov--Larkin
conductivity correction in the Cooper channel. This correction is
important in the vicinity of superconducting transition caused by
the inter-electron attraction~\cite{varlamov-larkin}.
However, in the case of the Coulomb repulsion, this correction is irrelevant,
both for the conductivity at $B=0$ and for the MC~\cite{Alt-Varlamov}.
\bibitem{foot-MTBto0} In the limit $\Omega_B\to 0$ the
Maki-Thompson term usually also dominates over the DoS-term
(both corrections give rise to a parabolic MC there). This holds as long as
$\lambda_c(T)[T\tau_\varphi]^2 \gtrsim 9 \zeta(3)/\pi^4 \simeq 0.11,$
the condition being
fulfilled in most relevant cases (since $T\tau_\varphi>1$,
while $\lambda_c(T)$ cannot be too low at experimentally accessible
temperatures).
\bibitem{footnote} In order
to match the result obtained from the expansion of the
digamma-function,
$\psi(1/2+1/\Omega_B\tau_\varphi(B))\simeq\psi(1/2)+\pi^2/2\Omega_B\tau_\varphi(B)$
with $\tau_\varphi(B)$ as defined in Eq.~(\ref{tfb}),
one should use a slightly different definition of
$\widetilde{\Omega}_B \equiv 2\Omega_B/\pi^2$ in
Eq.~(\ref{eq22a}). This is related to the fact that
Eq.~(\ref{eq22a}) was obtained by using the momentum integration
of the Cooperon instead of the summation over Landau levels. This
redefinition of $\Omega_B$ does not affect, however, the
logarithmic in $B$-terms appearing due to the interplay between
the WL and Coulomb interaction.
\end{thebibliography}
\end{document}